\documentclass[review,onefignum,onetabnum]{siamart190516}

\usepackage{lipsum}
\usepackage{amsfonts}
\usepackage{graphicx}
\usepackage{epstopdf}
\usepackage{algorithmic}
\usepackage{amssymb}

\usepackage{multirow,multicol,arydshln,colortbl}
\usepackage{algorithmic}

\ifpdf
  \DeclareGraphicsExtensions{.eps,.pdf,.png,.jpg}
\else
  \DeclareGraphicsExtensions{.eps}
\fi

\newsiamremark{remark}{Remark}
\newsiamremark{hypothesis}{Hypothesis}
\crefname{hypothesis}{Hypothesis}{Hypotheses}
\newsiamthm{claim}{Claim}
\newtheorem{assumption}{Assumption}
\newtheorem{example}{Example}

\headers{On identification of Boolean control networks}{B. Wang, and J.-e. Feng, and D. Z. Cheng}

\title{On identification of Boolean control networks\thanks{Submitted to the editors DATE.
\funding{This journal was supported by the National Natural Science Foundation of China (61773371, 61877036), and the Natural Science Fund of Shandong Province (ZR2019MF002, ZR2020QF117).}}}

\author{Biao Wang\thanks{School of Mathematics, Shandong University, Jinan, Shandong 250100, PR China,
  (\email{wangbiao@sdu.edu.cn}, \email{fengjune@sdu.edu.cn}).}
\and Jun-e Feng\footnotemark[2] \thanks{Corresponding author}
\and Daizhan Cheng\thanks{Key Laboratory of Systems and Control, Chinese
Academy of Sciences, Beijing 100190, PR China, (\email{dcheng@iss.ac.cn}).}
}

\usepackage{amsopn}

\ifpdf
\hypersetup{
  pdftitle={On identification of Boolean control networks},
  pdfauthor={B. Wang, J.-e. Feng, and D. Z. Cheng}
}
\fi

\begin{document}

\maketitle

\begin{abstract}
A new analytical framework consisting of two phenomena: single sample and multiple samples, is proposed to deal with the identification problem of Boolean control networks (BCNs) systematically and comprehensively.
Under this framework, the existing works on identification can be categorized as special cases of these two phenomena.
Several effective criteria for determining the identifiability and the corresponding identification algorithms are proposed.
Three important results are derived: (1) If a BN is observable, it is uniquely identifiable; (2) If a BCN is O1-observable, it is uniquely identifiable, where O1-observability is the most general form of the existing observability terms; (3) A BN or BCN may be identifiable, but not observable.
In addition, remarks present some challenging future research and contain a preliminary attempt about how to identify unobservable systems.
\end{abstract}



\section{Introduction}
Boolean network (BN), proposed by Kauffman in 1969 \cite{Kauffman1969}, is an ideal mathematical model of simulating the gene regulation networks.
It quantitates the interactions among genes within cells (or within a particular genome).
The expression, replication, transcription and other activities of genes can be directly reflected by system states and functions \cite{Kauffmanbook1993}.
BN has prompted many researchers to find and ask for similar models.
As a result, a large number of models were born.
For example, some genes continuously adjust the glucose consumption of cells and so provide the fuel by which they grow and multiply.
For analyzing such a biological system, Boolean control network (BCN) becomes a proper model \cite{Datta2004,Huang2000}.
One of the main tools for studying BNs and BCNs is called the semi-tensor product (STP) of matrices, which was proposed by Prof. Cheng \cite{Cheng2010BNlinear,Chengbook2011,Chengbook2012}.
Its basic idea is to describe the system behavior as a discrete time algebra form,
by which, some classical control ideas are incorporated into the analysis of BNs \cite{LiangJinBN2020,LiF-IBNS2012,LiRcompleteBN2012,Learning2018,ZhuJBN2015} and the control design of BCNs \cite{LiHLyapunov2019,LuJstab2018,WuYcontrol2018,Yuregular2019,ZhongJieTrank2019,QZhucontrol2019}, as people have seen in recent years.

Many wild animals carry multiple viruses that have no effect on the animals themselves, but may be both high contagious and deadly to human beings.
Antiviral immunity plays a key role against virus diseases.
Its research involves the pathologic manifestations, symptoms and detection technologies of viral disease, which is the major cause of network identification being currently an important topic.
Network identification aims to find the methods or algorithms for constructing the dynamics of systems.
For an unknown biological system or an environment where some viruses survive, only input-output data can be obtained, however, their changes can reflect some particular functions and features of a system.
Hence these data are directly used to build the model describing the original complicated network.
Some early results considered identification of the network transition mappings \cite{identification2000Akutsu,identification1998Liang,identification2006Nam,identification2005Pal}.
Under the framework of STP, the identification of BNs can be equivalently transformed into the identification of related structure matrices, which was proposed in \cite{ModelConstruction2011} and was extended to BCNs in \cite{IdentificationofBCN2011}.
\cite{IdentificationofBCN2011} noticed that, a BCN is identifiable if and only if it is controllable and O3-observable.
This O3-observability originates from one of five branching paths to the development of observability.
We list these five definitions of observability below.
\begin{definition}\label{Def}
A BCN is Oi-observable, {\rm($i=1,2,3,4,5$)}, if
\begin{itemize}
\item[{\rm(O1)}] {\rm \cite{OBCN1}} for any two distinct states $x(0)\neq \bar{x}(0)$, there exists an input sequence $(u(0), u(1), \ldots)$, such that the corresponding output sequences are distinct: $(y(0), y(1), $ $\ldots)\neq (\bar{y}(0), \bar{y}(1),\ldots)$;

\item[{\rm(O2)}] {\rm \cite{OBCN2}} for any a state $x(0)$ there exists an input sequence $(u(0)$, $u(1)$, $\ldots)$, such that for any $\bar{x}(0)\neq x(0)$, the corresponding output sequences are distinct: $(y(0), y(1), \ldots)\neq (\bar{y}(0), \bar{y}(1),\ldots)$;

\item[{\rm(O3)}] {\rm \cite{OBCN3}} there exists an input sequence $(u(0), u(1),\ldots)$, such that for any two distinct states $x(0)\neq \bar{x}(0)$, the corresponding output sequences are distinct: $(y(0), y(1), \ldots)\neq (\bar{y}(0), \bar{y}(1),\ldots)$;

\item[{\rm(O4)}] {\rm \cite{OBCN4}} for any two distinct states $x(0)\neq \bar{x}(0)$ and for any input sequence $(u(0),u(1),\ldots)$, the corresponding output sequences are distinct:
    $(y(0),y(1),$ $\ldots)$$\neq$ $(\bar{y}(0), \bar{y}(1),\ldots)$;

\item[{\rm(O5)}] {\rm \cite{OBCN5}} there exists an output-feedback loop $u(t)=f_t(y(t))$ $(u(t)=f(y(t))$ for static control), such that for any two distinct states $x(0)\neq \bar{x}(0)$, the corresponding output sequences are distinct: $(y(0), y(1), \ldots)\neq (\bar{y}(0), \bar{y}(1),\ldots)$.
\end{itemize}
\end{definition}

Most of the criteria and methods for judging the first four kinds of observability (O1-O4) are not ideal, some are sufficient conditions, some are too complex to apply.
\cite{OBCNZhang20161} proposed a unified approach based on finite automata to determine these four observabilities, and presented corresponding four necessary and sufficient conditions.
This automata approach is more suitable for BCNs with fewer state nodes and input nodes due to the high complexity of constructing deterministic finite automata.
\cite{OBCNCheng20162} concentrated on the most general observability (O1), and presented a matrix-based approach with lower complexity by STP.
Mathematically speaking, the requirement of input sequences used to recognize the initial state $x(0)$ gradually increases from O1-observability (the most general form) to O4-observability (the sharpest form). Hence, O4 $\Longrightarrow$ O3 $\Longrightarrow$ O2 $\Longrightarrow$ O1, which is shown as a relation diagram in \cite{OBCNZhang20161}.
In particular, determining O3-observability is NP-hard \cite{OBCN3}.
O5-observability that recognizes the initial state via output feedback, called output-feedback observability, was first proposed in \cite{OBCNGuoCCC2017}. This one is much sharper than the first two observabilities.
\cite{OBCN5} used paralled interconnected two identical BCNs to determine O5-observability, by converting the observability problem of the original BCN to the set reachability problem of the interconnected BCN.

As mentioned above, the identification of BCNs requires O3-observability.
A natural question arises: what about the most general form, O1-observability?
Motivated by that, we further develop the identification problem for BCNs in this paper.
Main contributions are summarized as follows:

(1) Three important theoretical results are obtained: (3a) A BN is uniquely identifiable if it is observable; (3b) A BCN is uniquely identifiable if it is O1-observable. It is worth pointing out that O1-observability is the most general one of the existing observability terms. (3c) A BN or BCN may be identifiable, but not observable.

(2) In combination with the phenomena in medical detection, we propose two new concepts: single sample and multiple samples to deal with the identification problem of BCNs.
Based on them, the identification problem is divided into four situations.
We point out that the existing works on identification are actually special cases of these four situations.

(3) By virtue of the observability property, we form a one-to-one correspondence between the state and the output sequence.
Then four simple criteria to determine the identifiability and four effective algorithms to construct the structure matrices are proposed.

The rest of the paper is organized as follows. Section II contains preliminary notations, fundamental definitions and problem formulation.
Section III presents main results on identification of BNs and BCNs, including several discriminant methods for the identification property, several identification algorithms to construct the structure matrices and illustrative examples. Remarks are given to show some challenging and interesting future research.
Finally, a table describes the relationships and comparisons of the results obtained in this paper, and Section IV concludes the paper.

\section{Preliminaries}

\subsection{Semi-tensor product}
This section gives some necessary preliminaries. More details can be referred to \cite{Chengbook2012}.
First, some notations are listed below:
\begin{itemize}
\item[$\bullet$] $\mathbb{N}=\{0,1,2,\ldots\}$: the natural number set.
\item[$\bullet$] $[a,b]_{\mathbb{N}}$: all the natural numbers from $a$ to $b$.
\item[$\bullet$] $\delta_n^i$: the $i$th column of the identity matrix $I_n$.
\item[$\bullet$] $\Delta_n:=\{\delta_n^1,\delta_n^2,\ldots,\delta_n^n\}$.
\item[$\bullet$] $\mathbf{1}_n:=\sum_{i=1}^n\delta_n^i$.
\item[$\bullet$] $[\delta_n^{i_1}~\delta_n^{i_2}~\cdots ~\delta_n^{i_m}]:=$ $\delta_n[i_1~i_2 ~\cdots~i_m]$.
\item[$\bullet$] $(\delta_n^{i_1},\delta_n^{i_2},\ldots,\delta_n^{i_m}):=$ $\delta_n(i_1,i_2,\ldots,i_m)$.
\item[$\bullet$] $Col_i(M)$: the $i$th column of matrix $M$.
\item[$\bullet$] $Col(M)$: the set of all columns of $M$.
\item[$\bullet$] $\mathcal{L}_{m\times n}$: $=\{M| M\in \mathbb{R}^{m\times n},Col(M)\subseteq \Delta_{m}\}$.
\item[$\bullet$] $[M]_{i,j}$: the $(i,j)$th entry of matrix $M$.
\item[$\bullet$] $M^{\mathrm{T}}$: the transpose of matrix $M$.
\item[$\bullet$] Kronecker product: $A\otimes B=([A]_{i,j}\times B)$.
\item[$\bullet$] K-R product: $A\ast B=C$, $Col_l(C)=Col_l(A)\otimes Col_l(B)$.

\end{itemize}

\begin{definition}{\rm\cite{Chengbook2012}}
The semi-tensor product (STP) of two matrices $A\in\mathbb{R}^{m\times n}$ and $B\in\mathbb{R}^{p\times q}$ is
\begin{equation*}
A\ltimes B=(A\otimes I_{\frac{s}{n}})(B\otimes I_{\frac{s}{p}}),
\end{equation*}
where $s$ is the least common multiple of $n$ and $p$.
\end{definition}

Obviously, the STP becomes the conventional matrix product if $n=p$. Hence the symbol $\ltimes$ is omitted in the sequel.

\begin{lemma}\label{Lem1}{\rm\cite{Chengbook2012}}
Let $f(x_1,\ldots,x_n)$ be a Boolean function, where $x_1,\ldots,x_n$ are Boolean variables.
Within the framework of vector form, $f$ can be converted into $f:\Delta_{2^n} \rightarrow \Delta_2$, and there exists a unique matrix $M_f\in \mathcal{L}_{2\times 2^n}$, called the structure matrix of $f$, such that
\begin{equation*}
f(x_1,\ldots,x_n)=M_f\ltimes x_1\ltimes \cdots\ltimes x_n.
\end{equation*}
\end{lemma}

Consider a BCN with $n$ state nodes, $m$ input nodes and $l$ output nodes as follows:
\begin{equation}\label{BCN1}
\left\{
\begin{array}{ll}
x_i(t+1)=f_i(u_1(t),\ldots,u_m(t),x_1(t),\ldots,x_n(t)),\\
y_j(t)=h_j(x_1(t),\ldots,x_n(t)),\\
i\in [1,n]_{\mathbb{N}},~j\in [1,l]_{\mathbb{N}},~t\in \mathbb{N},
\end{array}
\right.
\end{equation}
where $f_i:\mathcal{D}^{m+n}\rightarrow\mathcal{D}$ and $h_j:\mathcal{D}^{n}\rightarrow\mathcal{D}$ are logical functions, $x_i\in \mathcal{D}$, $u_j\in\mathcal{D}$ and $y_k\in\mathcal{D}$ are the state, input and output of the system, respectively.

From Lemma \ref{Lem1}, each logical function $f_i$ ($h_j$) has unique structure matrix $M_{f_i}$ ($M_{h_j}$), then BCN \eqref{BCN1} can be equivalently transformed into an algebraic form as follows \cite{Chengbook2012}:
\begin{equation}\label{BCN2}
\left\{
\begin{array}{ll}
x(t+1)=Fu(t)x(t),\\
y(t)=Hx(t),
\end{array}
\right.
\end{equation}
where $F=M_{f_1}\ast M_{f_2}\ast\cdots\ast M_{f_n}$, $H=M_{h_1}\ast M_{h_2}\ast\cdots\ast M_{h_l}$, $x(t)=\ltimes_{i=1}^nx_i(t)$, $u(t)=\ltimes_{j=1}^mu_j(t)$, $y(t)=\ltimes_{k=1}^ly_k(t)$.
This form is called the algebraic state space representation of \eqref{BCN1}.

\subsection{Problem statement}

The identification problem of BCN \eqref{BCN2} is to construct two structure matrices $F$ and $H$ via available data. Denote
\begin{align*}
U_i(p_i):=&\{u_i(t)\}_{t=0}^{p_i}=(u_i(0),u_i(1),u_i(2),\ldots,u_i(p_i)), \\
X_i(p_i):=&\{x_i(t)\}_{t=0}^{p_i}=(x_i(0),x_i(1),x_i(2),\ldots,x_i(p_i)), \\
Y_i(p_i):=&\{y_i(t)\}_{t=0}^{p_i}=(y_i(0),y_i(1),y_i(2),\ldots,y_i(p_i)),
\end{align*}
and
\begin{align*}
\{U_i(p_i)\}:=&\{u_i(0),u_i(1),u_i(2),\ldots,u_i(p_i)\}, \\
\{X_i(p_i)\}:=&\{x_i(0),x_i(1),x_i(2),\ldots,x_i(p_i)\}, \\
\{Y_i(p_i)\}:=&\{y_i(0),y_i(1),y_i(2),\ldots,y_i(p_i)\}.
\end{align*}

\begin{definition}
A BCN \eqref{BCN2} is said to be identifiable, if its two structure matrices $F$ and $H$ can be determined via available data: input data $U_1(p_1),U_2(p_2),\ldots,$ $U_k(p_k)$ and observed data $Y_1(p_1),Y_2(p_2),\ldots,Y_k(p_k)$.
\end{definition}

A coordinate transformation $\omega=Gx$ could convert \eqref{BCN2} into the following algebraic form:
\begin{align}\label{BCN3}
\left\{
\begin{array}{ll}
\omega(t+1)=GF(I_{2^m}\otimes G^{\mathrm{T}})u(t)\omega(t)=:\widehat{F}u(t)\omega(t), \\
y(t)=HG^{\mathrm{T}}\omega(t)=:\widehat{H}\omega(t).
\end{array}
\right.
\end{align}
Due to the arbitrariness of state recognition, \eqref{BCN2} and \eqref{BCN3} are considered to be identical in the same input-output data frame, so the set of all possible $(\widehat{F},\widehat{H})$ becomes the equivalence class of $(F,H)$.
A identifiable BCN is also said to be $uniquely$ $identifiable$ in the sense of equivalence.

\begin{assumption}\label{Assum1}
This paper assumes the available data is sufficient.
In other words, the input data and the observed data contain all possible situations which the system could generate.
\end{assumption}

Generally speaking, densely populated cities are good places for virus or infectious diseases, which could spread easily from person to person.
The Centers for Disease Control and Prevention can collect a large number of samples from different patients infected by the same pathogen.
Hence, Assumption \ref{Assum1} is reasonable and its implementation requires $multiple$ $samples$ from large numbers of patients (urine sample or blood sample or cheek swab), not a $single$ $sample$ from one patient, since a single sample may exhibit only part of characteristics of the virus.
Multiple samples mean that the observed data may be generated from different initial states, while, single sample means that the observed data is generated from some initial state.

On the basis of the statement above, the identification of BNs and BCNs can be divided into four cases:
\begin{itemize}
\item[Case 1]: the identification process of single sample in the BN records one group of output data $Y(p)$.

\item[Case 2]: the identification process of multiple samples in the BN records $k$ groups of output data $Y_1(p_1), Y_2(p_2), \ldots, Y_k(p_k)$.

\item[Case 3]: the identification process of single sample in the BCN records $r$ groups of input-output data $U_1(p_1),Y_1(p_1),U_2(p_2),Y_2(p_2),\ldots, U_r(p_r),Y_r(p_r)$.

\item[Case 4]: the identification process of multiple samples in the BCN records $rk$ groups of input-output data $U_1^i(p_1),Y_1^i(p_1),U_2^i(p_2),Y_2^i(p_2),\ldots, U_r^i(p_r),Y_r^i(p_r)$, $i\in [1,k]_{\mathbb{N}}$.

\end{itemize}
Both Cases 1 and 3 collect the blood sample from only one patient, while Cases 2 and 4 collect from $k$ patients.
Case 3 divides the blood sample into multiple portions ($r$ portions) for testing with a variety of reagents. That is to say, $r$ groups of input-output data are generated from the same initial state, i.e., $x_1(0)=x_2(0)=\cdots=x_r(0)$ (in Case 3). Similarly, $x_1^{i}(0)=x_2^i(0)=\cdots=x_r^i(0), i\in[1,k]_{\mathbb{N}}$ in Case 4.
%
%
%

\section{Identification of BNs and BCNs}

\subsection{Identification of BNs}
\cite{ModelConstruction2011} investigated the identification of the following BN:
\begin{align}\label{BN1}
\left\{
\begin{array}{ll}
x(t+1)=Fx(t),\\
y(t)=x(t),
\end{array}
\right.
\end{align}
in which the observed data is presented directly by the system state.
With a group of observed data $(x_1(0)=\delta_{2^n}^{i_0},x_1(1)=\delta_{2^n}^{i_1},\ldots)$, the $i_0$th column of $F$ can be identified as $Col_{i_0}(F)=\delta_{2^n}^{i_1}$ and hence the next result is obtained.

\begin{lemma}\label{Lem2}{\rm \cite{ModelConstruction2011}}
{\rm (}Multiple samples{\rm)} BN \eqref{BN1} is uniquely identifiable, if and only if the observed data contains all possible states:
\begin{align}
\{Y_1(p_1)\}\cup\{Y_2(p_2)\}\cup\cdots\cup\{Y_k(p_k)\}=\Delta_{2^n}.
\end{align}
\end{lemma}

It is noted that the observed data considered in \cite{ModelConstruction2011} may consist of several output sequences (i.e., multiple samples), which is reasonable because a system may contains multiple attractors and multiple attractors mean multiple state trajectories.
When the system state cannot be directly observed, BN \eqref{BN1} becomes
\begin{align}\label{BN2}
\left\{
\begin{array}{ll}
x(t+1)=Fx(t),\\
y(t)=Hx(t).
\end{array}
\right.
\end{align}
In the process of identifying this system, it is the most important to distinguish the states.

\begin{definition}
In BN \eqref{BN2}, a state pair $(x(0),\bar{x}(0))$, $x(0)\neq \bar{x}(0)$ is said to be distinguishable if the corresponding output sequences generated by them are distinct: $(y(0),y(1),\ldots)\neq (\bar{y}(0),\bar{y}(1),\ldots)$. \eqref{BN2} is said to be observable if any state pair is distinguishable.
\end{definition}

Observability means that $2^n$ distinct initial states generate $2^n$ distinct groups of observed data, i.e.,
\begin{align}
(H\delta_{2^n}^i,HF\delta_{2^n}^i,HF^{2}\delta_{2^n}^i,\ldots)\neq (H\delta_{2^n}^{i'},HF\delta_{2^n}^{i'},HF^{2}\delta_{2}^{i'},\ldots),~i\neq i'.
\end{align}
Since each state trajectory will fall into an attractor in $2^n$ steps, the subsequent state trajectory and output trajectory will repeat the previous data. Lemma 1 and Proposition 1 in \cite{OBCN4} show the following result.
\begin{proposition}\label{Pro3}{\rm\cite{OBCN4}}
BN \eqref{BN2} is observable if and only if for any $i\neq i'$,
\begin{align}\label{Pro3-1}
(H\delta_{2^n}^i,HF\delta_{2^n}^i,\ldots, HF^{2^n-1}\delta_{2^n}^i)\neq (H\delta_{2^n}^{i'},HF\delta_{2^n}^{i'},\ldots,HF^{2^n-1}\delta_{2}^{i'}).
\end{align}
\end{proposition}
We call $(H\delta_{2^n}^i,HF\delta_{2^n}^i,\ldots,HF^{2^n-1}\delta_{2^n}^i)$ the $effective$ $output$ $sequence$ of state $\delta_{2^n}^i$.
An effective output sequence corresponds to a state and its length is $2^n$ steps.
Under the case of Assumption \ref{Assum1}, if the system is observable, $2^n$ distinct effective output sequences can be found by searching and comparing all $2^n$-step output sequences from sufficient observed data.

Assume that the following $k$ groups of observed data are sufficient,
\begin{align}\label{Th1-data}
Y_j(T_j)=(y_j(0),y_j(1),\ldots,y_j(T_j)),~j\in [1,k]_{\mathbb{N}}.
\end{align}
Let $Y_s^j$ represent the $s$th $2^n$-step output sequence to show up in $Y_j(T_j)$:
\begin{align}\label{Th1-data1}
Y_s^j=&(y_j(s-1),y_j(s),\ldots,y_j(s+2^n-2)),~s\in [1,T_j']_{\mathbb{N}},
\end{align}
where $T_j'=T_j-2^n+2$.
Then by retrieval from \eqref{Th1-data}, an algorithm (Algorithm \ref{Alg:A}) to find $2^n$ distinct effective output sequences is established, and this algorithm names the $i$th effective output sequence that occurs in Algorithm \ref{Alg:A} as $Y_i$, $(i\in[1,2^n]_{\mathbb{N}})$.

\begin{algorithm}[H]
\caption{Retrieve all distinct effective output sequences}
\label{Alg:A}
\renewcommand{\algorithmicrequire}{\textbf{Input:}}
\renewcommand{\algorithmicensure}{\textbf{Output:}}
\begin{algorithmic}[1]
     \REQUIRE data \eqref{Th1-data}.
     \ENSURE $Y_1,Y_2,\ldots,Y_{2^n}$.
    \STATE{set $Y=\emptyset$ and $i=1$}
    \FOR{$j=1; j<k; j++$}
    \FOR{$s=1; s<T_j'; s++$}
        \IF {$Y_s^j\in Y$}
            \STATE break;
        \ELSE
            \STATE {$Y_i=Y_s^j$, $Y=Y\cup Y_i^j$, $i=i+1$;}
        \ENDIF
    \ENDFOR
    \ENDFOR
\end{algorithmic}
\end{algorithm}

\begin{theorem}\label{Th1}
{\rm(}Multiple samples{\rm)} BN \eqref{BN2} is uniquely identifiable if it is observable.
\end{theorem}
\begin{proof} From the analysis above, if \eqref{BN2} is observable, all distinct effective output sequences $Y_i,i\in[1,2^n]_{\mathbb{N}}$ can be obtained by Algorithm \ref{Alg:A} and enough observed data \eqref{Th1-data}.

Identify $Y_i$ as the effective output sequence of state $\delta_{2^n}^i$, $i\in[1,2^n]_{\mathbb{N}}$, then the following $k$ state sequences:
\begin{align}\label{Th1-data2}
X_j(T_j')=(x_j(0),x_j(1),\ldots,x_j(T_j')),~j\in[1,k]_{\mathbb{N}}
\end{align}
can be derived by
\begin{align}\label{Th1-data3}
x_j(t)=\delta_{2^n}^{i_t},~t\in[0,T_j'],~{\rm if}~(y_j(t),y_j(t+1),\ldots,y_j(t+2^n-1))=Y_{i_t}.
\end{align}
Combining the observed data \eqref{Th1-data} and the state data \eqref{Th1-data2}, $F$ and $H$ can be constructed by
\begin{align}\label{Th1-data4}
\left\{
\begin{array}{lll}
x_j(t+1)=Fx_j(t),~t\in[0,T_j'-1]_{\mathbb{N}},~j\in[1,k]_{\mathbb{N}},\\
H=[Y_1(0)~Y_2(0)~\cdots~Y_{2^n}(0)],
\end{array}
\right.
\end{align}
where $Y_i(0)$ represents the first element in $Y_i$, $i\in[1,2^n]_{\mathbb{N}}$.

Note that the appointment $\delta_{2^n}^i$ of the effective output sequence $Y_i$ can be changed in any order, due to the arbitrariness of state recognition.
Then another $(\widehat{F},\widehat{H})$ is derived, which is analogous to the case of \eqref{BCN3}.
One kind of order corresponds to one coordinate transformation $w=Gx$, and $(\widehat{F},\widehat{H})$ can be written as:
\begin{align}\label{BN3}
\left\{
\begin{array}{ll}
w(t+1)=GFG^{\mathrm{T}}w(t)=:\widehat{F}w(t),\\
y(t)=Hx(t)=HG^{\mathrm{T}}w(t)=:\widehat{H}w(t),
\end{array}
\right.
\end{align}
which implies both $(F,H)$ and $(\widehat{F},\widehat{H})$ belong to an equivalence class.
Hence, BN \eqref{BN2} is uniquely identifiable.
\end{proof}

On the basis of Theorem \ref{Th1}, an algorithm (Algorithm \ref{Alg:1}) to identify $(F,H)$ is established.

\begin{algorithm}
\caption{Identify $(F,H)$ for BN \eqref{BN2} (Deal with Case 2).}
\label{Alg:1}
{\textbf{Input:}} data \eqref{Th1-data}.

{\textbf{Output:}} $F, H$.

\begin{description}

\item[Step $1$]: Find all distinct effective output sequences by Algorithm \ref{Alg:A}.

\item[Step $2$]: For $j\in[1,k]_{\mathbb{N}}$ and $t\in[0,T_j']$, identify the state sequence \eqref{Th1-data2} by \eqref{Th1-data3}.

\item[Step $3$]: Construct $F$ and $H$ based on \eqref{Th1-data4}.
\end{description}
\end{algorithm}

\begin{example}\label{Eg1}
Consider a BN with $3$ state codes and $1$ output code, and assume there are two groups of observed data:
\begin{align}\label{Eg1-0}
\left\{
\begin{array}{ll}
Y_1(14)=\delta_2(2,1,1,2,2,2,1,2,2,2,1,2,2,2,1),\\
~~~~~~~~=(y_1(0),y_1(1),y_1(2),y_1(3),\ldots,y_1(14)),\\
Y_2(12)=\delta_2(1,2,1,2,1,2,1,2,1,2,1,2,1),\\
~~~~~~~~=(y_2(0),y_2(1),y_2(2),y_2(3),\ldots,y_2(12)).
\end{array}
\right.
\end{align}
\end{example}

Step~$1$: All distinct effective output sequences are
\begin{align}\label{Eg1-1}
\left\{
\begin{array}{llll}
Y_1=(y_1(0),y_1(1),\ldots,y_1(7))&=\delta_2(2,1,1,2,2,2,1,2), \\
Y_2=(y_1(1),y_1(2),\ldots,y_1(8))&=\delta_2(1,1,2,2,2,1,2,2), \\
Y_3=(y_1(2),y_1(3),\ldots,y_1(9))&=\delta_2(1,2,2,2,1,2,2,2), \\
Y_4=(y_1(3),y_1(4),\ldots,y_1(10))&=\delta_2(2,2,2,1,2,2,2,1), \\
Y_5=(y_1(4),y_1(5),\ldots,y_1(11))&=\delta_2(2,2,1,2,2,2,1,2), \\
Y_6=(y_1(5),y_1(6),\ldots,y_1(12))&=\delta_2(2,1,2,2,2,1,2,2), \\
Y_7=(y_2(0),y_2(1),\ldots,y_2(7))&=\delta_2(1,2,1,2,1,2,1,2), \\
Y_8=(y_2(1),y_2(2),\ldots,y_2(8))&=\delta_2(2,1,2,1,2,1,2,1).
\end{array}
\right.
\end{align}

Step~$2$: Two state sequences are identified as
\begin{align}\label{Eg1-2}
\left\{
\begin{array}{lll}
X_1(8)=(x_1(0),x_1(1),x_1(2),\ldots,x_1(8))\\
~~~~~~~=\delta_8(1,2,3,4,5,6,3,4,5),\\
X_2(5)=(x_2(0),x_2(1),x_2(2),x_2(3),x_2(4),x_2(5))\\
~~~~~~~=\delta_8(7,8,7,8,7,8).
\end{array}
\right.
\end{align}

Step~$3$: Based on~$x_1(t+1)=Fx_1(t)$, $t\in [0,8]_{\mathbb{N}}$, we get
\begin{align*}
x_1(1)=Fx_1(0)&\Rightarrow\delta_8^2=F\delta_8^1,x_1(4)=Fx_1(3)\Rightarrow\delta_8^5=F\delta_8^4,\\
x_1(2)=Fx_1(1)&\Rightarrow\delta_8^3=F\delta_8^2,x_1(5)=Fx_1(4)\Rightarrow\delta_8^6=F\delta_8^5,\\
x_1(3)=Fx_1(2)&\Rightarrow\delta_8^4=F\delta_8^3,x_1(6)=Fx_1(5)\Rightarrow\delta_8^3=F\delta_8^6.
\end{align*}
Similarly, from~$x_2(t+1)=Fx_2(t)$, $t\in [0,5]_{\mathbb{N}}$, we get
\begin{align*}
x_2(1)=Fx_2(0)&\Rightarrow\delta_8^8=F\delta_8^7,x_2(2)=Fx_2(1)\Rightarrow\delta_8^7=F\delta_8^8.
\end{align*}
It is clear that $F=\delta_8[2~3~4~5~6~3~8~7]$.
On the other hand,
\begin{align*}
H=&[Y_1(0)~Y_2(0)~\cdots~Y_{8}(0)] \\
 =&[y_1(0)~y_1(1)~y_1(2)~y_1(3)~y_1(4)~y_1(5)~y_2(0)~y_2(1)] \\
 =&\delta_2[2~1~1~2~2~2~1~2].
\end{align*}
To sum up, the system is identified as:
\begin{align}\label{Eg1-4}
\left\{
\begin{array}{ll}
x(t+1)=\delta_8[2~3~4~5~6~3~8~7]x(t),\\
y(t)=\delta_2[2~1~1~2~2~2~1~2]x(t).
\end{array}
\right.
\end{align}

If we identify $Y_7\sim\delta_8^1$, $Y_8\sim\delta_8^2$, $Y_1\sim\delta_8^3$, $Y_2\sim\delta_8^4$, $Y_3\sim\delta_8^5$, $Y_4\sim\delta_8^6$, $Y_5\sim\delta_8^7$ and $Y_6\sim\delta_8^8$, then the corresponding BN becomes:
\begin{align}\label{Eg1-5}
\left\{
\begin{array}{ll}
x(t+1)=\delta_8[2~1~4~5~6~7~8~5]x(t),\\
y(t)=\delta_2[1~2~2~1~1~2~2~2]x(t).
\end{array}
\right.
\end{align}
Although two systems \eqref{Eg1-4} and \eqref{Eg1-5} are derived from different appointments of effective output sequences, both of them belong to an equivalence class.

From this example, we can discern a special case: the observed multiple data are exactly $2^n$ distinct sequences $(y_i(0),y_i(1),\cdots,y_i(2^N-1))$, $i=1,2,\cdots,2^n$. The BN can also be identified by appropriately extending the length of the observed data (further collection adds one more element to each sequence).
In addition, Algorithms \ref{Alg:A} and \ref{Alg:1} also work with insufficient data, in which case, there exist some identical groups of observed data induced by some states that are indistinguishable, such that two structure matrices constructed are of low dimensions ($F\in \mathcal{L}_{2^n\times s}, H\in \mathcal{L}_{2^l\times s}, s < 2^n$).

In previous studies \cite{OBCN4,WBNAHS2020}, the effective output sequences are used to judge the observability, which is usually implemented by a tool called observability matrix:
\begin{align}\label{O0}
\mathcal{O}=[(PH)^{\mathrm{T}}~(PHF)^{\mathrm{T}}~\cdots ~(PHF^{2^n-1})^{\mathrm{T}}]^{\mathrm{T}},
\end{align}
where $P=[1~2~\cdots~2^l]$ is a row vector consisting of $[1,2^l]_{\mathbb{N}}$ and aims to extract the superscripts of all column vectors in $H$. For example, if $H=\delta_2[1~2~2~1]$, $PH=[1~2~2~1]$.
The effective output sequence of state $\delta_{2^n}^i$ is recorded in the $i$th column of $\mathcal{O}$:
$Col_i(\mathcal{O})=[PH\delta_{2^n}^i~PHF\delta_{2^n}^i~\cdots~$ $PHF^{2^n-1}\delta_{2^n}^i]^{\mathrm{T}}$.
Therefore, finding $2^n$ effective output sequences is equivalent to constructing the observability matrix.

\begin{example}
Recall Example \ref{Eg1}.
\end{example}

$Y_i$ is identified as the of state $\delta_{2^n}^i$. Hence, we have
\begin{small}
\begin{align*}
\mathcal{O}=&
\left[
\begin{array}{cccccccccc}
Py_1(0)&Py_1(1)&\cdots&Py_1(5) &Py_2(0)&Py_2(1)     \\
Py_1(1)&Py_1(2)&\cdots&Py_1(6) &Py_2(1)&Py_2(2)     \\
\vdots&\vdots&\ddots&\vdots &\vdots&\vdots  \\
Py_1(7)&Py_1(8)&\cdots&Py_1(12) &Py_2(7)&Py_2(8)
\end{array}
\right] \\
=&\left[
\begin{matrix}
2&1&1&2&2&2 &1&2     \\
1&1&2&2&2&1 &2&1     \\
1&2&2&2&1&2 &1&2  \\
2&2&2&1&2&2 &2&1  \\
2&2&1&2&2&2 &1&2  \\
2&1&2&2&2&1 &2&1  \\
1&2&2&2&1&2 &1&2  \\
2&2&2&1&2&2 &2&1
\end{matrix}
\right],
\end{align*}
\end{small}
which is the observability matrix of system \eqref{Eg1-1}.

\begin{proposition}\label{Pro1}
If $\mathcal{O}$ and $\widehat{\mathcal{O}}$ are the observability matrix of \eqref{BN2} and \eqref{BN3}, respectively, then
\begin{align*}
\mathcal{O}=\widehat{\mathcal{O}}G~{or}~\mathcal{O}G^{\mathrm{T}}=\widehat{\mathcal{O}}.
\end{align*}
\end{proposition}
\begin{proof}
From \eqref{O0}, it is clear that:
\begin{align}\label{O01}
\widehat{\mathcal{O}}=[(P\widehat{H})^{\mathrm{T}}~(P\widehat{H}\widehat{F})^{\mathrm{T}}~
\cdots~(P\widehat{H}\widehat{F}^{2^n-1})^{\mathrm{T}}]^{\mathrm{T}}.
\end{align}
Combining $\widehat{H}=HG^{\mathrm{T}}$ and $\widehat{F}=GFG^{\mathrm{T}}$, we get
\begin{align}
\widehat{\mathcal{O}}=&[(PHG^{\mathrm{T}})^{\mathrm{T}}~(PHG^{\mathrm{T}}GFG^{\mathrm{T}})^{\mathrm{T}}~
\cdots~(PHG^{\mathrm{T}}(GFG^{\mathrm{T}})^{2^n-1})^{\mathrm{T}}]^{\mathrm{T}}\notag \\
=&[(PHG^{\mathrm{T}})^{\mathrm{T}}~(PHFG^{\mathrm{T}})^{\mathrm{T}}~\cdots
~(PHF^{2^n-1}G^{\mathrm{T}})^{\mathrm{T}}]^{\mathrm{T}},    \notag \\
\widehat{\mathcal{O}}G=&[(PHG^{\mathrm{T}})^{\mathrm{T}}~(PHFG^{\mathrm{T}})^{\mathrm{T}}~\cdots
~(PHF^{2^n-1}G^{\mathrm{T}})^{\mathrm{T}}]^{\mathrm{T}}G   \notag \\
=&[(PHG^{\mathrm{T}}G)^{\mathrm{T}}~(PHFG^{\mathrm{T}}G)^{\mathrm{T}}~\cdots~(PHF^{2^n-1}G^{\mathrm{T}}G)^{\mathrm{T}}]^{\mathrm{T}}  \notag \\
=&[(PH)^{\mathrm{T}}~(PHF)^{\mathrm{T}}~\cdots ~(PHF^{2^n-1})^{\mathrm{T}}]^{\mathrm{T}}=\mathcal{O}. \notag
\end{align}
\end{proof}

Introducing the observability matrix $\mathcal{O}$ facilitates MATLAB programming.
After constructing such an observability matrix from observed data, $x(t)$ can be determined as $\delta_{2^n}^i$ if $[Py(t)~Py(t+1)~\cdots~Py(t+2^n-1)]^{\mathrm{T}}=Col_{i}(\mathcal{O})$.
Proposition \ref{Pro1} shows that the observability matrix of some system becomes that of another system by elementary column transformations, which further explains that different orders \eqref{Th1-data2} (or observability matrices) give rise to different systems $(F,H)$.
However, these systems belong to an equivalence class.

\begin{remark}\label{Rem1}
If BN \eqref{BN2} is identifiable, then is it observable?
Consider a BN with $F=\delta_4[3~3~4~4]$ and $H=\delta_2[1~1~2~1]$. Its state transition diagram is shown in Fig. \ref{fig:1}.
\begin{figure}[ht]
\begin{center}
\includegraphics[width=2.5in]{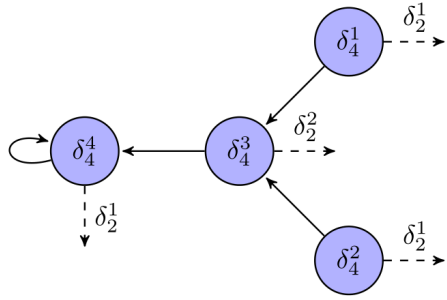}
\end{center}
\caption{The state transition diagram of an unobservable BN.}
\label{fig:1}
\end{figure}
This BN is unobservable since $\delta_4^1$ and $\delta_4^2$ have same effective output sequence.

When the system is unknown, with sufficient data $Y(8)=\delta_2(1,2,1,1,1,1,1,1,1)$ we can only find three distinct output sequences
\begin{align*}
Y_1=\delta_2(1,2,1,1),~Y_2=\delta_2(2,1,1,1),~Y_3=\delta_2(1,1,1,1).
\end{align*}
Then the corresponding state sequence can be determined as $X(5)=\delta_4(1,2,3,3,3,3)$.
If we know $Y_4=Y_1$, $x(1)=Fx(0)$ produces $\delta_4^2=F\delta_4^1$ and $\delta_4^2=F\delta_4^4$.
It follows that $F=\delta_4[2~3~3~2]$ and $H=\delta_2[1~2~1~1]$. This system and the original system coincide in the sense of equivalence.
This example shows that an unobservable system may be identifiable.
\end{remark}

\begin{remark}
To determine the observability, the dimensions of observability matrix $\mathcal{O}$ could be $(2^n-1)\times 2^n$ (see \cite{OBCN4,WBNAHS2020}).
That is to say, the condition \eqref{Pro3-1} in Proposition \ref{Pro3} can be improved as
\begin{align}
(H\delta_{2^n}^i,HF\delta_{2^n}^i,\ldots,HF^{2^n-2}\delta_{2^n}^i)\neq (H\delta_{2^n}^{i'},HF\delta_{2^n}^{i'},\ldots,HF^{2^n-2}\delta_{2^n}^{i'}),
\end{align}
which implies that the length of the effective output sequence can be further reduced to $(2^n-1)$ steps.
In a $2^n$-step output sequence $Y$, its first $(2^n-1)$ steps and last $(2^n-1)$ steps, respectively denoted by $f_e(Y)$ and $l_e(Y)$, reflect two effective output sequences, and hence they reflect two consecutive states.
Assume $Y_i$ and $Y_{i'}$ are two output sequences stemming from $\delta_{2^n}^i$ and $\delta_{2^n}^{i'}$, respectively.
If $l_e(Y_i)=f_e(Y_{i'})$, then state $\delta_{2^n}^{i'}$ follows state $\delta_{2^n}^i$, i.e., $\delta_{2^n}^{i'}=F\delta_{2^n}^i$.
So, comparing $Y_1,Y_2,\ldots,Y_{2^n}$ generated by Algorithm \ref{Alg:A}, we can directly construct $F$ and $H$.
For example, \eqref{Eg1-1} satisfies
\begin{align*}
&l_e(Y_1)=f_e(Y_2),~l_e(Y_2)=f_e(Y_3),~l_e(Y_3)=f_e(Y_4),~l_e(Y_4)=f_e(Y_5),\\
&l_e(Y_5)=f_e(Y_6),~l_e(Y_6)=f_e(Y_3),~l_e(Y_7)=f_e(Y_8),~l_e(Y_8)=f_e(Y_7).
\end{align*}
Then, $F\delta_8^1=\delta_8^2,F\delta_8^2=\delta_8^3,F\delta_8^3=\delta_8^4,F\delta_8^4=\delta_8^5, F\delta_8^5=\delta_8^6,F\delta_8^6=\delta_8^3,F\delta_8^7=\delta_8^8,F\delta_8^8=\delta_8^7$.
Obviously, this process is simpler than Algorithm \ref{Alg:1}.
However, extending this method to BCNs is not easy and requires consideration of the input sequence.
\end{remark}

\subsection{Identification of BCNs}

The identification problem of BCNs is complicated because more attractors will emerge as the control is embedded.
Considering BCN \eqref{BCN2}, the identification of $F\in \mathcal{L}_{2^{n}\times 2^{m+n}}$ needs the input-state data $(U_1(p_1), X_1(p_1), \ldots,U_k(p_k),X_k(p_k))$ that could cover all the possibilities of $\Delta_{2^{m+n}}$.
Then the identification problem is equivalent to checking
\begin{align}
\bigcup_{i=1}^k\{U_i(p_i)X_i(p_i)\}=\Delta_{2^{m+n}},
\end{align}
where $U_i(p_i)X_i(p_i)$ represents $(u_i(0)x_i(0),u_i(1)x_i(1),\ldots,u_i(p_i)x_i(p_i))$.

An ideal situation that only one group $(U_1(p_1), X_1(p_1))$ (not $k$ groups) is used to solve the identification problem of $F$, which was divided into two cases $H=I_{2^n}$ and $H\neq I_{2^n}$, and was analyzed in \cite{IdentificationofBCN2011}.
The observed data considered in \cite{IdentificationofBCN2011} can be seen as a special case of Case 3.
Next we slightly explain the method of \cite{IdentificationofBCN2011} and use $(U(p), X(p), Y(p))$ instead of $(U_1(p_1),X_1(p_1),Y_1(p_1))$ for simplicity.

Under a designable input sequence $U(p)$, the input-state pair $(u(t),x(t))$ to traverse all the possibilities of $\Delta_{2^{m+n}}$ can be guaranteed by controllability.
BCN \eqref{BCN2} is said to be $controllable$, if for any initial state $x(0)=x_0\in \Delta_{2^n}$ and any destination state $x_d\in \Delta_{2^n}$, there exists an input sequence $U(T)$, such that $x(T+1)=x_d$ \cite{OBCN2}.
When the system state is directly observable $(H=I_{2^n})$, the following lemma is immediate.
\begin{lemma}\label{Lem3}{\rm \cite{IdentificationofBCN2011}}
Let $H$ be the identity matrix $I_{2^n}$, then BCN \eqref{BCN2} is uniquely identifiable if and only if \eqref{BCN2} is controllable.
\end{lemma}

If a system is controllable, a single test sample could traverse all the possibilities.
Hence Lemma \ref{Lem3} is based on only one group of input-output data.
For the general case of $H\neq I_{2^n}$, the observability is involved again.

The O3-observability property of BCN \eqref{BCN2} can be viewed as the observability property of BN \eqref{BN2} to some extent.
Given an input sequence, a BCN is subject to a fixed evolutionary mechanism, similar to a BN.
Therefore, our main purpose is to find all distinct output sequences generated by an input sequence that makes \eqref{BCN2} O3-observable. 
\cite{IdentificationofBCN2011} designed an enough input sequence $U(T')$ which could determine the state sequence $X(T)$ $(T<T')$ by combining the corresponding output sequence $Y(T'+1)$, where $X(T)$ is the previous part of $X(T'+1)$ generated by $U(T')$.

\begin{lemma}\label{Lem4}{\rm \cite{IdentificationofBCN2011}}
BCN \eqref{BCN2} is uniquely identifiable from input-output data if and only if it is controllable and O3-observable.
\end{lemma}

Lemma \ref{Lem4} is also based on only one group of input-output data.
By controllability, there exists an input sequence $U(T')$, such that the system runs along the following state trajectory $X(T'+1)$ \cite{IdentificationofBCN2011}:
\begin{align}
&~~X(T'+1)=~(\underbrace{x(0),x(1),x(2),\ldots,x(T)}_{controllability}, \notag\\
&\underbrace{x(T+1),x(T+2),\ldots,\overset{x(0)}{\overset{\shortparallel}{x(t_{i_0}}}}_{controllability}\hspace{-1mm}\underbrace{), x(t_{i_0}+1),x(t_{i_0}+2),\ldots,x(t_{i_0}+p+1)}_{O3-observability},\notag\\
&\underbrace{x(t_{i_0}+p+2),x(t_{i_0}+p+3),\ldots,\overset{x(1)}{\overset{\shortparallel}{x(t_{i_1}}}}_{controllability}\hspace{-1mm}\underbrace{),x(t_{i_1}+1),x(t_{i_1}+2),\ldots,x(t_{i_1}+p+1)}_{O3-observability},
\notag \\
\label{XT-0}&\cdots, x(T'+1)),
\end{align}
where $T'+1=t_{i_T}+p$, and $X(T)$ satisfies
\begin{align}\label{XT-01}
\hspace{-0.2cm}
\{U(T)X(T)\}=\Delta_{2^{m+n}}.
\end{align}

To explain the purpose of driving the system along this state trajectory, we take the second and third rows as an example.
The state transition from $x(T+1)$ to $x(t_{i_0})=x(0)$ aims to find $x(0)$ by controllability, and that from $x(t_{i_0})$ to $x(t_{i_0}+p+1)$ aims to determine the value of $x(0)$ by O3-observability.
Continuing this process, BCN \eqref{BCN2} can be identified if these moments $t_{i_0}, t_{i_1},\ldots, t_{i_{T}}$ are in place.
Obviously, it is difficult to design such an input sequence $U(T')$ before the system is known, especially for the input sequence $(u(T+1),u(T+2),\ldots,u(T'))$ after time $T$, which involves timing accuracy.

Here we provide a method to identify $(F,H)$ with the help of multiple samples (Case 3), which is different from \cite{IdentificationofBCN2011} that used a single one.
An input sequence is called an O3-$test$ of BCN \eqref{BCN2} if it makes \eqref{BCN2} O3-observable.
From \eqref{XT-0}, we know that the input sequence $(u(T+1),u(T+2),\ldots,u(T'))$ aims to find and identify each state in $(x(0),x(1),\ldots,x(T))$.
Next, we simplify this process with the help of Case 3.

Assume that $(u_0',u_1',$ $\ldots,u_p')$ is an O3-test, and $(u_0,u_1,\ldots,u_T)$ satisfies
\begin{align}\label{Alg:2-1}
\{u(t)x(t)|u(t)=u_t,t\in[0,T]_{\mathbb{N}}\}=\Delta_{2^{m+n}}.
\end{align}
Then an O3-test could infiltrate the state at each time step by the following input sequences:
   \begin{align}\label{Alg:2-3}
   \left\{
   \begin{array}{lll}
   U_0(p)=(u_0',u_1',\ldots,u_p'),~~~~~~~~~~~~~~\\
   U_1(p+1)=(u_0,u_0',u_1',\ldots,u_p'), \\
   U_2(p+2)=(u_0,u_1,u_0',u_1',\ldots,u_p'),\\
       ~~~~~~\cdots\\
   U_{T+1}(p+T+1)=(u_0,u_1,\ldots,u_T,u_0',u_1',\ldots,u_p').
   \end{array}
   \right.
   \end{align}
Denote the corresponding observed data stemming from $x_0(0), x_1(0),\ldots, x_{T+1}(0)$ ($x_0(0)=x_1(0)=\cdots=x_{T+1}(0)$) as
    \begin{align}\label{Alg:2-4}
    \left\{
    \begin{array}{ll}
    Y_0(p+1)=(y_0(0),y_0(1),\ldots,y_0(p+1)), ~~~~~ \\
    Y_1(p+2)=(y_1(0),y_1(1),\ldots,y_1(p+2)), \\
    ~~~~~~\cdots \\
    Y_{T+1}(p+T+2)=(y_{T+1}(0),y_{T+1}(1),\ldots,y_{T+1}(p+T+2)).
    \end{array}
    \right.
    \end{align}
Since the system is O3-observable with respect to $(u_0',u_1',\ldots,u_p')$, we can find all distinct output sequences by retrieval from
    \begin{align}\label{Alg:2-5}
    Y^j=(y_j(j),y_j(j+1),\ldots,y_j(j+p+1)),~j\in [0,T+1]_{\mathbb{N}}.
    \end{align}
The corresponding retrieval algorithm (Algorithm \ref{Alg:B}) is given.
\begin{algorithm}[H]
\caption{Retrieve all distinct output sequences}
\label{Alg:B}
\renewcommand{\algorithmicrequire}{\textbf{Input:}}
\renewcommand{\algorithmicensure}{\textbf{Output:}}
\begin{algorithmic}[1]
     \REQUIRE data \eqref{Alg:2-4}.
     \ENSURE $Y_1,Y_2,\ldots,Y_{2^n}$.
    \STATE{set $Y=\emptyset$ and $i=1$}
    \FOR{$j=0; j<T+1; j++$}
        \IF {$Y^j\in Y$}
            \STATE continue;
        \ELSE
            \STATE {$Y_i=Y^j$, $Y=Y\cup \{Y_i\}$, $i=i+1$;}
        \ENDIF
    \ENDFOR
\end{algorithmic}
\end{algorithm}

Identify $Y_i$ as the output sequence of state $\delta_{2^n}^i$ with respect to $(u_0',u_1',\ldots,u_p')$, then BCN \eqref{BCN2} can be identified by input-output data \eqref{Alg:2-3} and \eqref{Alg:2-4}.
On the basis of the discussion above, we provide an algorithm (Algorithm \ref{Alg:2}) to identify $(F,H)$.

\begin{algorithm}
\caption{Identify $(F,H)$ for a controllable and O3-observable BCN \eqref{BCN2} (Deal with Case 3).}
\label{Alg:2}
{\textbf{Input:}} an O3-test $(u_0',u_1',\ldots,u_p')$, and $(u_0,u_1,\ldots,u_T)$ satisfying \eqref{Alg:2-1}.

{\textbf{Output:}} $F, H$.
	
\begin{description}
\item[Step $1$]: Construct input sequences \eqref{Alg:2-3} and record the observed data \eqref{Alg:2-4}.

\item[Step $2$]: Find all distinct output sequences $Y_1,Y_2,\ldots,Y_{2^n}$ by Algorithm \ref{Alg:B}.

\item[Step $3$]: Identify the state sequence $X_{T+1}(T+1)=(x_{T+1}(0),x_{T+1}(1),\ldots,x_{T+1}(T+1))$
    by
    \begin{align*}
    x_{T+1}(t)=\delta_{2^n}^{i_t},~t\in[0,T+1]_{\mathbb{N}},~{\rm if}~(y_{t}(t),y_{t}(t+1),\ldots,y_{t}(t+p+1))=Y_{i_t}.
    \end{align*}

\item[Step $4$]: Construct $F$ and $H$ based on
    \begin{align}\label{Alg:2-6}
    \left\{
    \begin{array}{lll}
    x_{T+1}(t+1)=Fu_tx_{T+1}(t),~t\in [0,T], \\
    H=[Y_1(0)~Y_2(0)~\cdots~Y_{2^n}(0)],
    \end{array}
    \right.
    \end{align}
    where $Y_i(0)$ represents the first element in $Y_i$, $i\in[1,2^n]_{\mathbb{N}}$.
\end{description}
\end{algorithm}

The following example originates from literature \cite{IdentificationofBCN2011}, in which an identification process with single input sequence is shown.
We use it to show how to use Algorithm \ref{Alg:2}.
\begin{example}\label{Eg2}
{\rm\cite{IdentificationofBCN2011}}
Consider an O3-observable and controllable BCN \eqref{BCN2} with
\begin{align}\label{Eg2-1}
\left\{
\begin{array}{ll}
F=&\delta_4[2~4~1~1~2~3~2~2], \\
H=&\delta_2[2~1~1~2].
\end{array}
\right.
\end{align}
\end{example}

Fix the initial state $x(0)=\delta_4^1$ and set
\begin{align*}
\left\{
\begin{array}{ll}
(u_0,u_1,\ldots,u_{10})=\delta_2(1, 1, 1, 2, 2, 1, 1, 1, 2, 2, 2), \\
(u_0',u_1')=\delta_2(1,1).
\end{array}
\right.
\end{align*}
Case 3 divides the test sample into enough portions to be tested $(x_0(0)=x_1(0)=\cdots=x_r(0)=\delta_4^1)$.

Step 1: 12 groups of input sequence are constructed and are infiltrated to 12 test samples: $x_k(0)=\delta_4^1, k\in[0,11]$. Then the corresponding output sequences are
\begin{align*}
Y_0(2)     =&\delta_2(\underline{2,1,2}),\\
Y_1(3)     =&\delta_2(2,\underline{1,2,2}),\\
Y_2(4)     =&\delta_2(2,1,\underline{2,2,1}),    \\
Y_3(5)     =&\delta_2(2,1,2,2,1,2),\\
Y_4(6)     =&\delta_2(2,1,2,2,1,2,2),\\
Y_5(7)     =&\delta_2(2,1,2,2,1,\underline{1,2,1})\\
Y_6(8)     =&\delta_2(2,1,2,2,1,1,2,1,2), \\
Y_7(9)     =&\delta_2(2,1,2,2,1,1,2,1,2,2),\\
Y_8(10)    =&\delta_2(2,1,2,2,1,1,2,1,2,2,1),  \\
Y_9(11)    =&\delta_2(2,1,2,2,1,1,2,1,2,1,2,2), \\
Y_{10}(12) =&\delta_2(2,1,2,2,1,1,2,1,2,1,1,2,1), \\
Y_{11}(13) =&\delta_2(2,1,2,2,1,1,2,1,2,1,1,1,2,2).
\end{align*}

Step 2: All distinct output sequences generated by $(u_0',u_1')$ are
\begin{align*}
Y_1=&(y_0(0),y_0(1),y_0(2))=\delta_2(2,1,2),\\
Y_2=&(y_1(1),y_1(2),y_1(3))=\delta_2(1,2,2),\\
Y_3=&(y_2(2),y_2(3),y_2(4))=\delta_2(2,2,1),\\
Y_4=&(y_5(5),y_5(6),y_5(7))=\delta_2(1,2,1),
\end{align*}
which are the last three elements of $Y_0(2)$, $Y_1(3)$, $Y_2(4)$ and $Y_5(7)$, respectively.

Step 3: Since
\begin{align*}
&(y_0(0),y_0(1),y_0(2))=Y_1,~(y_1(1),y_1(2),y_1(3))=Y_2,\\
&~~~~~~~~~~~~~~~~~~~\cdots~~~~~~~~~~~~~~~~~~~~~~~~\cdots\\
&(y_{10}(10),y_{10}(11),y_{10}(12))=Y_4,~(y_{11}(11),y_{11}(12),y_{11}(13))=Y_2,
\end{align*}
one gets the state sequence
\begin{align*}
X_{11}(11)=&(x_{11}(0),x_{11}(1),\ldots,x_{11}(11)) \\
          =&\delta_4(1, 2, 3, 1, 2, 4, 1, 2, 3, 2, 4, 2).
\end{align*}

Step 4: Based on \eqref{Alg:2-6}, this BCN is identified as
\begin{align}\label{Eg2-2}
\left\{
\begin{array}{ll}
x(t+1)=\delta_4[2~3~1~1~2~4~2~2]u(t)x(t), \\
y(t)=\delta_2[2~1~2~1]x(t).
\end{array}
\right.
\end{align}

In the sense of O3-observability, the observability matrix of BCN \eqref{BCN2} is analogous to the observability matrix of BN \eqref{BN2}, and can be written as:
\begin{align}
\mathcal{O}=
\left[
\begin{array}{llll}
PH \\
PHFu_0'\\
PHFu_1'Fu_0'\\
~~~\vdots\\
PHFu_p'\cdots Fu_1'Fu_0'
\end{array}
\right]_{(p+2)\times 2^n}
\end{align}
which consists of $2^n$ distinct output sequences stemming from $2^n$ initial states under the input sequence $(u_0',u_1',\ldots,u_p')$. For example, the observability matrix $\mathcal{O}$ of \eqref{Eg2-2} is
\begin{align}
\mathcal{O}=
\left[
\begin{array}{ccccc}
2&1&2&1\\
1&2&2&2\\
2&2&1&1
\end{array}
\right],
\end{align}
where the $i$th column of $\mathcal{O}$ corresponds with $Y_i$, $i\in[1,4]_{\mathbb{N}}$.

From the discussion above, one sees that, the identification problem is transformed into three conditions:
\begin{description}
\item[1)]The state trajectory of the system covers the entire state space;

\item[2)]There exists a method to distinguish all states;

\item[3)]There exist methods to implement 1) and 2).
\end{description}
These three conditions can be guaranteed by controllability, observability, and Assumption \ref{Assum1}, respectively.
Under the case of Assumption \ref{Assum1}, we can find $2^n$ distinct effective output sequences, or $2^n$ distinct output sequences generated by an O3-test, (or say the observability matrix $\mathcal{O}$), based on available data.
If such output sequences are found or the observability matrix is constructed, the state sequence will be determined, and then the identification problem will be solved.

Case 3 relaxes the limitation on the number of input sequences and output sequences.
Although $(T+2)$ input sequences are used in Algorithm \ref{Alg:2}, it can be implemented in biological experiments by cloning or segmentation, like diagnosis of disease in blood samples.
Inspired by the idea of multiple input sequences, next Lemma \ref{Lem4} will be extended to the most general form O1-observability.

Considering an O1-observable BCN \eqref{BCN2}, there exist $N:=2^{2n-1}-2^{n-1}$ input sequences $U_{(i,i')}(p)$, $1\leq i\neq i' \leq 2^n$, such that $\delta_{2^n}^i$ and $\delta_{2^n}^{i'}$ can be distinguishable by $U_{(i,i')}(p)$. Arrange these input sequences as
\begin{align}\label{O1-test}
U_{s(i,i')}(p)=(u_{s(i,i')}(0),\ldots,u_{s(i,i')}(p))=U_{(i,i')}(p),
\end{align}
where the function $s(i,i')$ is defined by the following rule:
\begin{align}
s(i,i')=\left\{
\begin{array}{ll}
i'-i,~i=1, \\
\sum_{j=1}^{i-1}(n-j)+i'-i,~i\geq 2.
\end{array}
\right.
\end{align}
We call \eqref{O1-test} an O1-$test$ of BCN \eqref{BCN2}.
Let $Y_{i,s}=(y_i^s(0),y_i^s(1),\ldots,y_i^s(p+1))$ represent the output sequence stemming from $\delta_{2^n}^i$ with respect to $U_s(p)$, where $y_i^s(t)=HFu_s(t-1)\cdots Fu_s(1)Fu_s(0)\delta_{2^n}^i$.
Construct data arrays $D_i$, $i\in [1,2^n]_{\mathbb{N}}$ as
\begin{align}\label{Di}
D_i=(Y_{i,1}, Y_{i,2}, \ldots, Y_{i,N}).
\end{align}
Then for any distinct state $\delta_{2^n}^i$ and $\delta_{2^n}^{i'}$, we have
$Y_{i,s(i,i')}\neq Y_{i',s(i,i')}$ and $D_i\neq D_{i'}$.

\begin{proposition}\label{Pro2}
BCN \eqref{BCN2} is O1-observable if and only if there exist $2^n$ data arrays constructed by \eqref{Di} satisfying
$D_{i}\neq D_{i'},~1\leq i\neq i'\leq 2^{n}$.
\end{proposition}

Proposition \ref{Pro2} shows that $2^n$ distinct states correspond to $2^n$ distinct data arrays in the sense of O1-observability.
Hence, analogous to the method used in O3-observability, by infiltrating O1-test \eqref{O1-test} into $x(0),x(1),\ldots, x(T+1)$ at each time step, we construct input sequences $U_0^s,U_1^s,\ldots,U_{T+1}^s$, $s\in [1,N]_{\mathbb{N}}$ as
   \begin{align}\label{Alg:3-2}
   \left\{
   \begin{array}{lll}
   U_0^s=(u_s(0),u_s(1),\ldots,u_s(p)),~~~~~\\
   U_1^s=(u_0,u_s(0),u_s(1),\ldots,u_s(p)), \\
   U_2^s=(u_0,u_1,u_s(0),u_s(1),\ldots,u_s(p)),\\
    ~~~~\cdots\\
   U_{T+1}^s=(u_0,u_1,\ldots,u_T,u_s(0),u_s(1),\ldots,u_s(p)),
   \end{array}
   \right.
   \end{align}
and denote the corresponding observed data as,
    \begin{align}\label{Alg:3-3}
    \left\{
    \begin{array}{lll}
    Y_0^s=(y_0^s(0),y_0^s(1),\ldots,y_0^s(p+1)),~~~~~\\
    Y_1^s=(y_1^s(0),y_1^s(1),\ldots,y_1^s(p+2)),\\
    Y_2^s=(y_2^s(0),y_2^s(1),\ldots,y_2^s(p+3)),\\
        ~~~~\cdots \\
    Y_{T+1}^s=(y_{T+1}^s(0),y_{T+1}^s(1),\ldots,y_{T+1}^s(T+p+2)),
    \end{array}
    \right.
    \end{align}
where $(u_0,u_1,\ldots,u_T)$ satisfies \eqref{Alg:2-1}.

All data arrays generated by O1-test are recorded in
    \begin{align}\label{Alg:3-1}
    D_s^j=(y_j^s(j),y_j^s(j+1),\ldots,y_j^s(j+p+1)),~j\in [0,T+1]_{\mathbb{N}}.
    \end{align}
According the following retrieval algorithm (Algorithm \ref{Alg:C}), we can find all distinct data arrays $D_i$, $i\in[1,2^n]_{\mathbb{N}}$.
\begin{algorithm}[H]
\caption{Retrieve all distinct data arrays}
\label{Alg:C}
\renewcommand{\algorithmicrequire}{\textbf{Input:}}
\renewcommand{\algorithmicensure}{\textbf{Output:}}
\begin{algorithmic}[1]
     \REQUIRE data \eqref{Alg:3-1}.
     \ENSURE $D_1,D_2,\ldots,D_{2^n}$.
    \STATE{set $D=\emptyset$ and $i=1$}
    \FOR{$j=0; j<T+1; j++$}
        \STATE{$D^j=(D_1^j,D_2^j,\ldots,D^j_{N})$}
        \IF {$D^j\in D$}
            \STATE continue;
        \ELSE
            \STATE {$D_i=D^j$, $D=D\cup \{D_i\}$, $i=i+1$;}
        \ENDIF
    \ENDFOR
\end{algorithmic}
\end{algorithm}
Identify $O_i$ as the data array stemming from $\delta_{2^n}^i$, then the state sequence $(x(0),$ $x(1),\ldots,x(T+1))$ can be identified by
\begin{align}\label{Alg:3-4}
x(t)=\delta_{2^n}^{i_t},~{\rm if}~D^t=D_{i_t}.
\end{align}
Combining $(u_0,u_1,\ldots,u_T)$, $(x(0),x(1),\ldots, x(T+1))$ and $(y_{T+1}^1(0),\ldots, y_{T+1}^1(T+1))$, $F$ and $H$ can be easily obtained.

\begin{theorem}\label{Th2} {\rm(}Single sample{\rm)}
BCN \eqref{BCN2} is uniquely identifiable if it is controllable and O1-observable.
\end{theorem}

On the basis of the discussion above, we give an algorithm (Algorithm \ref{Alg:3}) to identify BCN \eqref{BCN2} which is O1-observable.
\begin{algorithm}
\caption{Identify $(F,H)$ for a controllable and O1-observable BCN \eqref{BCN2} (Deal with Case 3).}
\label{Alg:3}
{\textbf{Input:}} O1-test: $U_1(p), U_2(p),\ldots,U_{N}(p)$, and $(u_0,u_1,\ldots,u_T)$ satisfying \eqref{O1-test}.

{\textbf{Output:}} $F, H$.
\begin{description}

\item[Step $1$]: Construct input sequences \eqref{Alg:3-2} and record the observed data \eqref{Alg:3-3}.

\item[Step $2$]: Find all distinct data arrays by Algorithm \ref{Alg:C}.

\item[Step $3$]: Identify the state sequence $X(T+1)=(x(0),x(1),\ldots,x(T+1))$
    by \eqref{Alg:3-4}.
\item[Step $4$]: Construct $F$ and $H$ based on $X(T+1)$, $(u_0,u_1,\ldots,u_T)$, $(y_{T+1}^1(0),\ldots,$ $y_{T+1}^1(T+1))$ and
    \begin{align}\label{Alg:3-5}
    \left\{
    \begin{array}{lll}
    x(t+1)=Fu_tx(t),  \\
    y_{T+1}^1(t)=Hx(t), \\
    t\in [0,T].
    \end{array}
    \right.
    \end{align}

\end{description}
\end{algorithm}

Algorithm \ref{Alg:3} is not a complex process, for the method used in \cite{IdentificationofBCN2011}, should be relatively straightforward.
The core of the identification problem is the complete determination of state sequence $X(T+1)$.
Compared with \cite{IdentificationofBCN2011} (see \eqref{XT-0}), the method proposed here considers only the first $T+1$ states and inputs, and avoids the design of subsequent inputs $u(T+1),u(T+2),\ldots$.
Note that the determination of the state sequence depends on the appointment of $D_i$. Hence $(F,H)$ generated by Algorithm \ref{Alg:3} belongs to the equivalence class of the original system.

\begin{remark}
One idea for proving the necessity is to employ Lemma \ref{Lem4} and a fact that O3-observability implies O1-observability.
Logically, this is misleading, because, Theorem \ref{Th2} is based on multiple input sequences, and Lemma \ref{Lem4} is based on single input sequence.
Hence, we have
identifiability $\Longleftarrow$ controllability $+$ O1-observability $\Longleftarrow$ controllability $+$ O3-observability.
\end{remark}

As predicted, the implement of multiple input sequences (Case 3) allows more numerous systems to be identified.
Next we consider whether the conditions of Theorem \ref{Th2} be further relaxed. The answer is yes.
In Case 4, a large number of samples from different patients are collected and each sample is divided into multiple portions.
If the state sequences $X_j^i(p)=(x_j^i(0),x_j^i(1),\ldots,x_j^i(p))$ and the input sequences $U_j^i(p)=(u_j^i(0),u_j^i(1),\ldots,u_j^i(p))$ cover all the possibilities of $\Delta_{2^{m+n}}$, i.e.,
\begin{align}
\bigcup_{i=1,j=1}^{k,r}\{U_j^i(p)X_j^i(p)\}=\Delta_{2^{m+n}},
\end{align}
then by infiltrating the O1-test to all states $x_j^i(t), i\in[1,k]_{\mathbb{N}}, j\in[1,r]_{\mathbb{N}}, t\in[0,p]_{\mathbb{N}}$, the identification problem can be solved.

\begin{definition}
Consider BCN \eqref{BCN2} with a set of initial states $P_0$ and a set of destination states $P_d$.

{\rm i)} $P_d$ is set-$P_0$ controllable, if for any $x_d\in P_d$, there exist a state $x(0)=x_0\in P_0$ and an input sequence $U(T)$, such that $x(T+1)=x_d$;

{\rm ii)} \eqref{BCN2} is set-$P_0$ controllable, if $\Delta_{2^n}$ is set-$P_0$ controllable.

{\rm iii)} A state $x_0$ is controllable, if \eqref{BCN2} is set-$P_0$ controllable and $P_0$ is a singleton set $P_0=\{x_0\}$.

\end{definition}

\begin{theorem}\label{Th3} {\rm(}Multiple samples{\rm)}
Suppose the set of initial states is $P_0$, then BCN \eqref{BCN2} is uniquely identifiable if it is set-$P_0$ controllable and O1-observable.
\end{theorem}

\begin{proof}
Without loss of generality, assume $P_0=\{\delta_{2^n}^1,\delta_{2^n}^2,\ldots,\delta_{2^n}^k\}$, and the state space is split into $\Delta_{2^n}=P_d^1\cup P_d^2\cup \cdots \cup P_d^k$, such that $P_d^a$ is set-$\{\delta_{2^n}^a\}$ controllable, $a\in[1,k]_{\mathbb{N}}$. ($P_0$ may contain more (than $k$) states and we only use those which make the above condition true.)

Consider state $\delta_{2^n}^a$ and set $P_d^a$, if $P_d^a=\delta_{2^n}\{d_1^a,d_2^a,\ldots,d_{\theta_a}^a\}$, then for any state $\delta_{2^n}^i\in P_d^a$, there exists an input sequence
\begin{align}
U_{a\rightarrow i}(p_i^a)=(u_{a\rightarrow i}(0),u_{a\rightarrow i}(1),\ldots,u_{a\rightarrow i}(p_i^a)),
\end{align}
such that $Fu_{a\rightarrow i}(p_i^a)\cdots Fu_{a\rightarrow i}(1)Fu_{a\rightarrow i}(0)\delta_{2^n}^a=\delta_{2^n}^i$.

Construct an input sequence as
\begin{align}
\label{Th3-2} U^{j}_{a\rightarrow i}(p_i^a+1)=&(u_{a\rightarrow i}^j(0),u_{a\rightarrow i}^j(1),\ldots,u_{a\rightarrow i}^j(p_i^a),u_{a\rightarrow i}^j(p_i^a+1)) \notag \\
=&:(u_{a\rightarrow i}(0),u_{a\rightarrow i}(1),\ldots,u_{a\rightarrow i}(p_i^a),\delta_{2^m}^j),
\end{align}
under which, the state sequence stemming from $\delta_{2^n}^a$ becomes
\begin{align}
X^{j}_{a\rightarrow i}(p_i^a+2)=&(x_{a\rightarrow i}^j(0),x_{a\rightarrow i}^j(1),\ldots,x_{a\rightarrow i}^j(p_i^a+1),x_{a\rightarrow i}^j(p_i^a+2)) \notag \\
  =&:(\delta_{2^n}^a,x_{a\rightarrow i}(1),\ldots,\delta_{2^n}^i,x_{a\rightarrow i}^j(p_i^a+2)),
\end{align}
then we have
\begin{align}
\delta_{2^m}^j\delta_{2^n}^{i}=u_{a\rightarrow i}^j(p_i^a+1)x_{a\rightarrow i}^j(p_i^a+1)\in \{U^{j}_{a\rightarrow i}(p_i+1)X^j_{a\rightarrow i}(p_i+1)\}.
\end{align}

It follows that
\begin{align}
\Delta_{2^{m+n}}=&\bigcup_{j=1}^{2^m}\big\{\delta_{2^m}^j\delta_{2^n}^i| \delta_{2^n}^i\in \Delta_{2^n}\big\} \notag \\
=&\bigcup_{j=1}^{2^m}\bigcup_{a=1}^{k}\big\{\delta_{2^m}^j\delta_{2^n}^i| \delta_{2^n}^i\in P_d^a\}\big\} \notag \\
\subseteq& \bigcup_{j=1}^{2^m}\bigcup_{a=1}^{k}\big\{U^{j}_{a\rightarrow i}(p_i+1)X^j_{a\rightarrow i}(p_i+1)|i\in\{d_1^a,d_2^a,\ldots,d_{\theta_a}^a\}\big\}\notag \\
=&\bigcup_{j=1}^{2^m}\bigcup_{a=1}^{k}\bigcup_{i=1}^{2^n}\big\{U^{j}_{a\rightarrow i}(p_i+1)X^j_{a\rightarrow i}(p_i+1)\big\},
\end{align}
which implies that enough input-state data could cover $\Delta_{2^{m+n}}$.

By infiltrating the O1-test \eqref{O1-test} to all states in $X^{j}_{a\rightarrow i}(p_i^a+2)$, BCN \eqref{BCN2} can be identified. 

\end{proof}

\begin{corollary}\label{Cor3-1}{\rm(}Multiple samples{\rm)}
Suppose BCN \eqref{BCN2} is O1-observable and the initial state set is $P_0$. If $P_d$ is the maximum set which is set-$P_0$ controllable, then $|P_d|$ states can be identified.
\end{corollary}

Theorem \ref{Th3} and Corollary \ref{Cor3-1} show that, the maximum level of identifying BCN \eqref{BCN2} depends on the relationship of all possible states stemming from the initial state set.
If the controllability is not available, the identification problem can still be well done with the same method.
In densely populated cities, the Centers for Disease Control and Prevention can collect a large number of samples, which most likely contain all possible initial states.
That is to say, all possible test samples endowed with $2^n$ distinct initial states are collected.
In Case 4, assume that $U_s$ \eqref{O1-test} is an O1-test of the system, and $x_1^i(0)=x_2^i(0)=\cdots=x_r^i(0)=\delta_{2^n}^i$, $i\in[1,2^n]_{\mathbb{N}}$.
Construct input sequences
\begin{align}\label{Alg:4-1}
\left\{
\begin{array}{lll}
U_s^i(p)       =(u_s^i(0),u_s^i(1),\ldots,u_s^i(p))=U_s,~s\in[1,N]_{\mathbb{N}}, \\
U_{jN+s}^i(p+1)  =(u_{jN+s}^i(0),u_{jN+s}^i(1),\ldots,u_{jN+s}^i(p+1))\\
~~~~~~~~~~~~~~~~~~=(\delta_{2^m}^j,U_s),~j\in[1,2^m]_{\mathbb{N}},~s\in[1,N]_{\mathbb{N}},
\end{array}
\right.
\end{align}
then we have
   \begin{align}
   \Delta_{2^{m+n}}=&\bigcup_{j=1}^{2^m}\bigcup_{i=1}^{2^n}\{\delta_{2^m}^j\delta_{2^n}^i\} \notag \\
                   =&\bigcup_{j=1}^{2^m}\bigcup_{i=1}^{2^n}\big\{u_{jN+s}^i(0)x_{jN+s}^i(0)\big\} \notag \\
                   \subseteq &\bigcup_{j=1}^{2^m}\bigcup_{i=1}^{2^n}\big\{U_{jN+s}^i(p) X_{jN+s}^i(p)\big\}.
   \end{align}
Then the system can be easily identified analogous to Theorem \ref{Th3}.
Hence we have the following result under the case of Assumption 1.
\begin{theorem}\label{Th4} {\rm(}Multiple samples{\rm)}
BCN \eqref{BCN2} is uniquely identifiable if it is O1-observable.
\end{theorem}

On the basis of the analysis above, the corresponding identification algorithm (Algorithm \ref{Alg:4}) can be established.
\begin{algorithm}
\caption{Identify an O1-observable BCN \eqref{BCN2} (Deal with Case 4).}
\label{Alg:4}
{\textbf{Input:}} O1-test \eqref{O1-test}: $U_1(p), U_2(p),\ldots,U_{N}(p)$.

{\textbf{Output:}} $F,H$.

\begin{description}

\item[Step $1$]: Construct input sequences \eqref{Alg:4-1} and record the observed data
   \begin{align}\label{Alg:4-2}
   \left\{
   \begin{array}{lll}
   Y_s^i(p+1)=(y_s^i(0),y_s^i(1),\ldots,y_s^i(p+1)),~s\in[1,N]_{\mathbb{N}}, \\
Y_{jN+s}^i(p+2)  =(Y_{jN+s}^i(0),\ldots,Y_{jN+s}^i(p+2)),~j\in[1,2^m]_{\mathbb{N}},~s\in[1,N]_{\mathbb{N}}.
\end{array}
\right.
\end{align}

\item[Step $2$]: Find all distinct data arrays by Algorithm \ref{Alg:D}.

\item[Step $3$]: Identify the states
    $x_1^i(0)$, $x_{jN+1}^i(1)$, $j\in [1,2^m]_{\mathbb{N}}, i\in[1,2^n]_{\mathbb{N}}$, by
    \begin{align}
    \left\{
    \begin{array}{lll}
    x_1^i(0)=\delta_{2^n}^{a_i},~{\rm if}~(Y_1^i(p+1),\ldots,Y_N^i(p+1))=D_{a_i}, \\
    x_{Nj+1}^i(1)=\delta_{2^n}^{a_i},~{\rm if}~(\overline{Y}_{jN+1}^i,\ldots,\overline{Y}_{jN+s}^i)=D_{a_i},
    \end{array}
    \right.
    \end{align}
    where $\overline{Y}_{jN+s}^i=(y_{jN+s}^i(1),\ldots,y_{jN+s}^i(p+2))$.
\item[Step $4$]: Construct $F$ and $H$ based on
    \begin{align}\label{Alg:4-4}
    \left\{
    \begin{array}{lll}
    x_{Nj+1}^i(1)=F\delta_{2^m}^jx_{Nj+1}^i(0)=F\delta_{2^m}^jx_{1}^i(0),  \\
    y_{1}^i(0)=Hx_{1}^i(0).
    \end{array}
    \right.
    \end{align}

\end{description}

\end{algorithm}

\begin{algorithm}[H]
\caption{Retrieve all distinct data arrays}
\label{Alg:D}
\renewcommand{\algorithmicrequire}{\textbf{Input:}}
\renewcommand{\algorithmicensure}{\textbf{Output:}}
\begin{algorithmic}[1]
     \REQUIRE data \eqref{Alg:4-2}.
     \ENSURE $D_1,D_2,\ldots,D_{2^n}$.
    \STATE{set $D=\emptyset$ and $a=1$}
    \FOR{$i=0; i<2^n; i++$}
    \FOR{$j=0; j<2^n; j++$}
        \IF {$j=0$}
        \STATE{$D_s^j=Y_s^i(p+1)$}
        \ELSE
           \STATE{$D_s^j=(y_{jN+s}^i(1),\ldots,y_{jN+s}^i(p+2))$}
        \ENDIF
        \STATE{$D^j=(D_1^j,D_2^j,\ldots,D^j_{N})$}
        \IF {$D^j\in D$}
            \STATE continue;
        \ELSE
            \STATE {$D_a=D^j$, $D=D\cup D_a$, $a=a+1$;}
        \ENDIF
    \ENDFOR
    \ENDFOR
\end{algorithmic}
\end{algorithm}

\begin{remark}
There are two ways to construct the structure matrix $H$. One is based on the state sequence and the output sequence, like \eqref{Alg:3-5} and \eqref{Alg:4-4} (in Algorithms \ref{Alg:3} and \ref{Alg:4}), the other is based on the first elements of the effective output sequences or the output sequences generated by O3-test or the data arrays generated by O1-test, like \eqref{Th1-data4} and \eqref{Alg:2-6} (in Algorithms \ref{Alg:1} and \ref{Alg:2}).
\end{remark}

\begin{remark}
It is important to point out that, the input sequence required for identification is assumed to be known, whether in {\rm\cite{IdentificationofBCN2011}} {\rm(}used O3-observability{\rm)} or our paper {\rm(}O1-observability{\rm)}.
For unknown systems, different input sequences are used to implement Algorithm \ref{Alg:4} (Algorithm \ref{Alg:2}, Algorithm \ref{Alg:3}) until the system is fully identified.
The former {\rm(}O3-observability{\rm)} covers an NP-hard problem{\rm \cite{OBCN3}}, which has been mentioned in the introduction part. So does our method relax, in a sense, the difficulty of dealing with the identification problem? In other words, is the difficulty of finding an O1-test also NP-hard?
\end{remark}

\begin{example}\label{Eg4}
Consider the reduced model for the lac operon in the bacterium Escherichia coli.
This BCN has three state nodes $\{x_1,x_2,x_3\}$ and three input nodes $\{u_1,u_2,u_3\}$.
$x_1$: lac mRNA, $x_2$: lactose in high concentration, $x_3$: lactose in medium concentration, $u_1$: extracellular glucose, $u_2$: high extracellular lactose, and $u_3$: medium extracellular lactose.
The dynamics of this system can be written as \cite{lacoperon2011}
\begin{align}\label{Eg4-1}
\left\{
\begin{array}{ll}
x_1(t+1)=\neg u_1(t)\wedge (x_2(t)\vee x_3(t)), \\
x_2(t+1)=\neg u_1(t)\wedge u_2(t) \wedge x_1(t), \\
x_3(t+1)=\neg u_1(t)\wedge (u_2(t)\vee (u_3(t)\wedge x_1(t))).
\end{array}
\right.
\end{align}
This BCN is O1-observable when the outputs are considered as \cite{SetcontrolCheng2018}
\begin{align}\label{Eg4-2}
\left\{
\begin{array}{ll}
y_1(t)=x_1(t)\vee \neg x_2(t)\vee x_3(t),\\
y_2(t)=\neg x_1(t)\vee x_2(t) \wedge \neg x_3(t),\\
y_3(t)=\neg x_1(t)\wedge \neg x_2(t)\vee x_3(t).
\end{array}
\right.
\end{align}
\end{example}
Its algebraic form is
\begin{align}\label{Eg4-3}
\left\{
\begin{array}{ll}
F=\delta_8[8~8~8~8~8~8~8~8 ~8~8~8~8~8~8~8~8 \\
~~~~~~~~~8~8~8~8~8~8~8~8 ~8~8~8~8~8~8~8~8 \\
~~~~~~~~~1~1~1~5~3~3~3~7 ~1~1~1~5~3~3~3~7 \\
~~~~~~~~~3~3~3~7~4~4~4~8 ~4~4~4~8~4~4~4~8],\\
H=\delta_8[8~6~3~6~5~6~7~6].
\end{array}
\right.
\end{align}

Here we analyze the identification problem of this system. Fix the initial state $x_1^i(0)=\cdots=x_r^i(0)=\delta_8^i$, $i\in[1,8]_{\mathbb{N}}$ and choose the following O1-test
\begin{align}\label{Eg4-O1-test}
U_s=\left\{
\begin{array}{lll}
(\delta_8^5),~s\in S,\\
(\delta_8^1),~s\in S^c,
\end{array}
\right.
\end{align}
where $S=\{9,10,11,12,13,20,21,22,27\}$ and $S^c=[1,28]_{\mathbb{N}}\setminus \{9,10,11,12,13,20,21,$ $22,27\}$.

Step 1: Construct input sequences
\begin{align*}
\left\{
\begin{array}{lll}
U_s^i(0)       =(u_s^i(0))=U_s,~s\in[1,28]_{\mathbb{N}}, \\
U_{28j+s}^i(1)  =(u_{28j+s}^i(0),u_{28j+s}^i(1))=(\delta_8^j,U_s),~j\in[1,8]_{\mathbb{N}}~s\in[1,28]_{\mathbb{N}},
\end{array}
\right.
\end{align*}
then the output sequences stemming from states $x_1^1(0),x_2^1(0),\ldots,x_r^1(0)$ are
\begin{align*}
\left\{
\begin{array}{lll}
Y_s^1(1)   =\delta_8(\underline{8,6}),~s\in S^c,\\
Y_s^1(1)   =\delta_8(\underline{8,8}),~s\in S,\\
Y_{28+s}^1(2) =Y_{56+s}^1(2)=Y_{84+s}^1(2) =Y_{112+s}^1(2) =\delta_8(8,\underline{6,6}),~s\in S^c,   \\
Y_{28+s}^1(2) =Y_{56+s}^1(2)=Y_{84+s}^1(2) =Y_{112+s}^1(2) =\delta_8(8,\underline{6,7}),~s\in S,   \\
Y_{140+s}^1(2) =Y_{168+s}^1(2) =\delta_8(8,{8,6}),~s\in S^c,   \\
Y_{140+s}^1(2) =Y_{168+s}^1(2) =\delta_8(8,{8,8}),~s\in S,   \\
Y_{196+s}^1(2) =\delta_8(8,\underline{3,6}),~s\in S^c,   \\
Y_{196+s}^1(2) =\delta_8(8,\underline{3,8}),~s\in S,   \\
Y_{224+s}^1(2) =\delta_8(8,\underline{6,6}),~s\in S^c,   \\
Y_{224+s}^1(2) =\delta_8(8,\underline{6,5}),~s\in S,
\end{array}
\right.
\end{align*}
and others are shown in Appendix \ref{appA}.


Step 2: All distinct data arrays generated by the O1-test \eqref{Eg4-O1-test} are
\begin{align*}
D_1=& (D_1^1,D_2^1,\ldots,D_{28}^1),~\left\{
\begin{array}{lll}
D_s^1=\delta_8(8,6),~s\in S^c,\\
D_s^1=\delta_8(8,8),~s\in S,
\end{array}
\right. \\
D_2=& (D_1^2,D_2^2,\ldots,D_{28}^2),~\left\{
\begin{array}{lll}
D_s^2=\delta_8(6,6),~s\in S^c,\\
D_s^2=\delta_8(6,7),~s\in S,
\end{array}
\right. \\
&\cdots \\
D_8=& (D_1^8,D_2^8,\ldots,D_{28}^8),~\left\{
\begin{array}{lll}
D_s^8=\delta_8(6,6),~s\in S^c,\\
D_s^8=\delta_8(6,3),~s\in S,
\end{array}
\right.
\end{align*}
which are the last two elements of $Y_s^1(1)$, $Y_{28+s}^1(2)$, $Y_{196+s}^1(2)$, $Y_{224+s}^1(2)$, $Y_{s}^2(2)$, $Y_{196+s}^4(2)$, $Y_{s}^5(1)$ and $Y_{s}^6(1)$, respectively.

Step 3: Identify $D_i$ as the data array stemming from $\delta_8^i$, $i\in[1,8]_{\mathbb{N}}$, then from Step 2, we have
\begin{align*}
x_{28+1}^1(1)=Fu_{28+1}^1(0)x_{28+1}^1(0)\Rightarrow &~ \delta_8^2=F\delta_8^1\delta_8^1, \\
x_{56+1}^1(1)=Fu_{56+1}^1(0)x_{56+1}^1(0)\Rightarrow &~ \delta_8^2=F\delta_8^2\delta_8^1, \\
x_{84+1}^1(1)=Fu_{84+1}^1(0)x_{84+1}^1(0)\Rightarrow &~ \delta_8^2=F\delta_8^3\delta_8^1, \\
x_{112+1}^1(1)=Fu_{112+1}^1(0)x_{112+1}^1(0)\Rightarrow &~ \delta_8^2=F\delta_8^4\delta_8^1, \\
x_{140+1}^1(1)=Fu_{140+1}^1(0)x_{140+1}^1(0)\Rightarrow &~ \delta_8^1=F\delta_8^5\delta_8^1, \\
x_{168+1}^1(1)=Fu_{168+1}^1(0)x_{168+1}^1(0)\Rightarrow &~ \delta_8^1=F\delta_8^6\delta_8^1, \\
x_{196+1}^1(1)=Fu_{196+1}^1(0)x_{196+1}^1(0)\Rightarrow &~ \delta_8^3=F\delta_8^7\delta_8^1, \\
x_{224+1}^1(1)=Fu_{224+1}^1(0)x_{224+1}^1(0)\Rightarrow &~ \delta_8^4=F\delta_8^8\delta_8^1, \\
\cdots &
\end{align*}
and
\begin{align*}
&y_{1}^1(0)=Hx_{1}^1(0)\Rightarrow~ \delta_8^8=H\delta_8^1,~y_{1}^2(0)=Hx_{1}^2(0)\Rightarrow ~ \delta_8^6=H\delta_8^5, \\
&y_{1}^3(0)=Hx_{1}^3(0)\Rightarrow~ \delta_8^3=H\delta_8^3,~y_{1}^4(0)=Hx_{1}^4(0)\Rightarrow ~ \delta_8^6=H\delta_8^4, \\
&y_{1}^5(0)=Hx_{1}^5(0)\Rightarrow~ \delta_8^5=H\delta_8^7,~y_{1}^6(0)=Hx_{1}^6(0)\Rightarrow ~ \delta_8^6=H\delta_8^8, \\
&y_{1}^7(0)=Hx_{1}^7(0)\Rightarrow~ \delta_8^7=H\delta_8^6,~y_{1}^8(0)=Hx_{1}^8(0)\Rightarrow~ \delta_8^6=H\delta_8^2.
\end{align*}
This BCN therefore is identified as
\begin{align}\label{Eg4-4}
\left\{
\begin{array}{ll}
F=\delta_8[2~2~2~2~2~2~2~2 ~2~2~2~2~2~2~2~2 \\
 ~~~~~~~~~2~2~2~2~2~2~2~2 ~2~2~2~2~2~2~2~2 \\
 ~~~~~~~~~1~6~1~7~1~3~3~3 ~1~6~1~7~1~3~3~3 \\
 ~~~~~~~~~3~2~3~6~3~4~4~4 ~4~2~4~2~4~4~4~4],\\
H=\delta_8[8~6~3~6~6~7~5~6].
\end{array}
\right.
\end{align}

\begin{remark}
A BN may be identifiable but not observable, as mentioned in Remark \ref{Rem1}.
Similar to the method in BNs (see Remark \ref{Rem1}), it is still possible to identify an unobservable BCN.
Therefore, observability is a stronger property than identifiability in BNs and BCNs.
\end{remark}

\begin{remark}
About the identification algorithm in the Matlab programming, the storage method and storage space of the identification standard affect the execution time of the algorithm, because the programming involves the storage and retrieval of data.
Hence, it is vital to plan for storage space needs at the beginning of the design phase.
Compared with $2^n$ distinct effective output sequences or $2^n$ distinct output sequences generated by an O3-test or the corresponding observability matrix, $2^n$ distinct data arrays generated by an O1-test require more storage space. How to adjust the storage method and how to reduce the storage space are two challenging and interesting topics.
\end{remark}

Up to now, we have provided several ways to deal with different situations.
The comparison of them is needed and is shown in Table I, where Case 3' considers one group of input-output data, a special case of Case 3.

\begin{table*}[!htbp]
\centering
\begin{tabular}{c | c | c | c |c }
\hline
Approach & System & Case & Condition 1 & Condition 2 \\
\hline
Lemma \ref{Lem2} \cite{ModelConstruction2011} & BN & Case 2&  & $H=I_{2^l}$ \\
\cdashline{1-5}[1pt/2pt]
Theorem \ref{Th1} & \multirow{2}{*}{BN} & \multirow{2}{*}{Case 2}  &  & \multirow{2}{*}{Observability} \\
Algorithm \ref{Alg:1}  &     &  &    &   \\
\cdashline{1-5}[1pt/2pt]
Lemma \ref{Lem3}\cite{IdentificationofBCN2011} & BCN & Case 3 & controllability & $H=I_{2^l}$ \\
\cdashline{1-5}[1pt/2pt]
Lemma \ref{Lem4}\cite{IdentificationofBCN2011} & \multirow{2}{*}{BCN} & Case $3'$ & \multirow{2}{*}{controllability} & \multirow{2}{*}{O3-observability}  \\
Algorithm \ref{Alg:2} &     & Case 3   &  &   \\
\cdashline{1-5}[1pt/2pt]
Theorem \ref{Th2} &   \multirow{2}{*}{BCN}   & \multirow{2}{*}{Case 3}   &\multirow{2}{*}{controllability} & \multirow{2}{*}{O1-observability}  \\
Algorithm \ref{Alg:3}   &   &  &   & \\
\cdashline{1-5}[1pt/2pt]
Theorem \ref{Th3} &  BCN  & Case 4   & Set controllability & O1-observability   \\
\cdashline{1-5}[1pt/2pt]
Theorem \ref{Th4} &  \multirow{2}{*}{BCN}  & \multirow{2}{*}{Case4}   &     & \multirow{2}{*}{O1-observability} \\
Algorithm \ref{Alg:4} &   &  &   & \\
\hline
\end{tabular}
\caption{A comparison table of various methods}
\end{table*}

\section{Conclusions}\label{sec:conclusions}

In this paper, we systematically explored the identification problem of BNs and BCNs, gained new cognition.
Based on the practical application, we built a new analytical framework that consists of single sample and multiple samples, and then divided the identification problem into four situations.
Four simple criteria were proposed for determining the identifiability of BNs and BCNs, and the corresponding identification algorithms were provided to identify related structure matrices.
It is worth noting that these algorithms are easy to implement by MATLAB.
Under this analytical framework, we found three novel and important results:
(1) A BN is uniquely identifiable if it is observable; (2) A BCN is uniquely identifiable if it is O1-observable;
(3) The necessity of (1) or (2) does not hold.
At last, we presented a table to reveal the relationships of the proposed results, which could be used to further analyze the relationships of identifiability, controllability, O1-observability and O3-observability, as a foreshadowing.

The authors believe that this new analytical framework is a very powerful explanation tool.
On the basis of this paper, there are several natural and interesting problems remaining for further study. For example: (1) If a BN or BCN is unobservable, some states produce same output sequence, which implies that some columns in the structure matrices $F$ and $H$ are indeterminate. A very natural question is whether an unobservable system can be identified, or, how to identify an unobservable system?
(2) The assumption (Assumption \ref{Assum1}) that the input-output data is sufficient is a highlight of this paper.
How to generalize this idea to other systems, like singular BNs or switched BCNs?

\appendix
\section{The observed data of Step~1 in Example \ref{Eg4}}\label{appA}
The output sequences stemming from states $x_1^2(0),x_2^2(0),\ldots,x_r^2(0)$ are
\begin{align*}
\left\{
\begin{array}{lll}
Y_s^2(1)  =\delta_8(\underline{6,6}),~s\in S^c,\\
Y_s^2(1)  =\delta_8(\underline{6,8}),~s\in S,\\
Y_{28+s}^2(2) =Y_{56+s}^2(2)=Y_{84+s}^2(2) =Y_{112+s}^2(2) =\delta_8(6,{6,6}),~s\in S^c,   \\
Y_{28+s}^2(2) =Y_{56+s}^2(2)=Y_{84+s}^2(2) =Y_{112+s}^2(2) =\delta_8(6,{6,7}),~s\in S,   \\
Y_{140+s}^2(2) =Y_{168+s}^2(2) =\delta_8(6,{8,6}),~s\in S^c,   \\
Y_{140+s}^2(2) =Y_{168+s}^2(2) =\delta_8(6,{8,8}),~s\in S,   \\
Y_{196+s}^2(2) =\delta_8(6,{3,6}),~s\in S^c,   \\
Y_{196+s}^2(2) =\delta_8(6,{3,8}),~s\in S,   \\
Y_{224+s}^2(2) =\delta_8(6,{6,6}),~s\in S^c,   \\
Y_{224+s}^2(2) =\delta_8(6,{6,5}),~s\in S.
\end{array}
\right.
\end{align*}
The output sequences stemming from states $x_1^3(0),x_2^3(0),\ldots,x_r^3(0)$ are
\begin{align*}
\left\{
\begin{array}{lll}
Y_s^3(1)   =\delta_8({3,6}),~s\in S^c,\\
Y_s^3(1)   =\delta_8({3,8}),~s\in S,\\
Y_{28+s}^3(2) =Y_{56+s}^3(2)=Y_{84+s}^2(2) =Y_{112+s}^3(2) =\delta_8(3,{6,6}),~s\in S^c,   \\
Y_{28+s}^3(2) =Y_{56+s}^3(2)=Y_{84+s}^2(2) =Y_{112+s}^3(2) =\delta_8(3,{6,7}),~s\in S,   \\
Y_{140+s}^3(2) =Y_{168+s}^3(2) =\delta_8(3,{8,6}),~s\in S^c,   \\
Y_{140+s}^3(2) =Y_{168+s}^3(2) =\delta_8(3,{8,8}),~s\in S,   \\
Y_{196+s}^3(2) =\delta_8(3,{3,6}),~s\in S^c,   \\
Y_{196+s}^3(2) =\delta_8(3,{3,8}),~s\in S,   \\
Y_{224+s}^3(2) =\delta_8(3,{6,6}),~s\in S^c,   \\
Y_{224+s}^3(2) =\delta_8(3,{6,5}),~s\in S.
\end{array}
\right.
\end{align*}
The output sequences stemming from states $x_1^4(0),x_2^4(0),\ldots,x_r^4(0)$ are
\begin{align*}
\left\{
\begin{array}{lll}
Y_s^4(1)   =\delta_8({6,6}),~s\in S^c,\\
Y_s^4(1)   =\delta_8({6,5}),~s\in S,\\
Y_{28+s}^4(2) =Y_{56+s}^4(2) =Y_{84+s}^4(2) \\
~~~~~~~~~~~=Y_{112+s}^4(2) =Y_{224+s}^4(2) =\delta_8(6,{6,6}),~s\in S^c,   \\
Y_{28+s}^4(2) =Y_{56+s}^4(2) =Y_{84+s}^4(2) \\
~~~~~~~~~~~=Y_{112+s}^4(2) =Y_{224+s}^4(2) =\delta_8(6,{6,7}),~s\in S,   \\
Y_{140+s}^4(2) =Y_{168+s}^4(2) =\delta_8(6,{5,6}),~s\in S^c,   \\
Y_{140+s}^4(2) =Y_{168+s}^4(2) =\delta_8(6,{5,3}),~s\in S,   \\
Y_{196+s}^4(2) =\delta_8(6,\underline{7,6}),~s\in S^c,   \\
Y_{196+s}^4(2) =\delta_8(6,\underline{7,3}),~s\in S.
\end{array}
\right.
\end{align*}
The output sequences stemming from states $x_1^5(0),x_2^5(0),\ldots,x_r^5(0)$ are
\begin{align*}
\left\{
\begin{array}{lll}
Y_s^5(1)   =\delta_8(\underline{5,6}),~s\in S^c,\\
Y_s^5(1)   =\delta_8(\underline{5,3}),~s\in S,\\
Y_{28+s}^5(2) =Y_{56+s}^5(2)=Y_{84+s}^5(2) =Y_{112+s}^5(2) =\delta_8(5,{6,6}),~s\in S^c,   \\
Y_{28+s}^5(2) =Y_{56+s}^5(2)=Y_{84+s}^5(2) =Y_{112+s}^5(2) =\delta_8(5,{6,7}),~s\in S,   \\
Y_{140+s}^5(2)=Y_{168+s}^5(2) = \delta_8(5,{3,6}),~s\in S^c,  \\
Y_{140+s}^5(2)=Y_{168+s}^5(2) = \delta_8(5,{3,8}),~s\in S,  \\
Y_{196+s}^5(2) =Y_{224+s}^5(2) =\delta_8(5,{6,6}),~s\in S^c,   \\
Y_{196+s}^5(2) =Y_{224+s}^5(2) =\delta_8(5,{6,5}),~s\in S.
\end{array}
\right.
\end{align*}
The output sequences stemming from states $x_1^6(0),x_2^6(0),\ldots,x_r^6(0)$ are
\begin{align*}
\left\{
\begin{array}{lll}
Y_s^6(1)   =\delta_8(\underline{6,6}),~s\in S^c,\\
Y_s^6(1)   =\delta_8(\underline{6,3}),~s\in S,\\
Y_{28+s}^6(2) =Y_{56+s}^6(2)=Y_{84+s}^6(2) =Y_{112+s}^3(2) =\delta_8(6,{6,6}),~s\in S^c,   \\
Y_{28+s}^6(2) =Y_{56+s}^6(2)=Y_{84+s}^6(2) =Y_{112+s}^3(2) =\delta_8(6,{6,7}),~s\in S,   \\
Y_{140+s}^6(2)=Y_{168+s}^6(2) = \delta_8(6,{3,6}),~s\in S^c,  \\
Y_{140+s}^6(2)=Y_{168+s}^6(2) = \delta_8(6,{3,8}),~s\in S,  \\
Y_{196+s}^6(2) =Y_{224+s}^6(2) =\delta_8(6,{6,6}),~s\in S^c,   \\
Y_{196+s}^6(2) =Y_{224+s}^6(2) =\delta_8(6,{6,5}),~s\in S.
\end{array}
\right.
\end{align*}
The output sequences stemming from states $x_1^7(0),x_2^7(0),\ldots,x_r^7(0)$ are
\begin{align*}
\left\{
\begin{array}{lll}
Y_s^7(1)   =\delta_8({7,6}),~s\in S^c,\\
Y_s^7(1)   =\delta_8({7,3}),~s\in S,\\
Y_{28+s}^7(2) =Y_{56+s}^7(2)=Y_{84+s}^7(2) =Y_{112+s}^7(2) =\delta_8(7,{6,6}),~s\in S^c,   \\
Y_{28+s}^7(2) =Y_{56+s}^7(2)=Y_{84+s}^7(2) =Y_{112+s}^7(2) =\delta_8(7,{6,7}),~s\in S,   \\
Y_{140+s}^7(2)=Y_{168+s}^7(2) = \delta_8(7,{3,6}),~s\in S^c,  \\
Y_{140+s}^7(2)=Y_{168+s}^7(2) = \delta_8(7,{3,8}),~s\in S, \\
Y_{196+s}^7(2) =Y_{224+s}^7(2) =\delta_8(7,{6,6}),~s\in S^c,   \\
Y_{196+s}^7(2) =Y_{224+s}^7(2) =\delta_8(7,{6,5}),~s\in S.
\end{array}
\right.
\end{align*}
The output sequences stemming from states $x_1^8(0),x_2^8(0),\ldots,x_r^8(0)$ are
\begin{align*}
\left\{
\begin{array}{lll}
Y_s^8(1)   =\delta_8({6,6}),~s\in S^c,\\
Y_s^8(1)   =\delta_8({6,7}),~s\in S,\\
Y_{28+s}^8(2) =Y_{56+s}^8(2)  =Y_{84+s}^8(2)  =Y_{112+s}^8(2) \\
~~~~~~~~~~~   =Y_{196+s}^8(2) =Y_{224+s}^8(2) =\delta_8(6,{6,6}),~s\in S^c,   \\
Y_{28+s}^8(2) =Y_{56+s}^8(2)  =Y_{84+s}^8(2)  =Y_{112+s}^8(2)\\
~~~~~~~~~~~   =Y_{196+s}^8(2) =Y_{224+s}^8(2) =\delta_8(6,{6,7}),~s\in S,   \\
Y_{140+s}^8(2) =Y_{168+s}^8(2) =\delta_8(6,{7,6}),~s\in S^c,   \\
Y_{140+s}^8(2) =Y_{168+s}^8(2) =\delta_8(6,{7,3}),~s\in S.
\end{array}
\right.
\end{align*}


\bibliographystyle{siamplain}
\bibliography{BiaoWang}
\end{document}